\documentclass[titlepage,chapternotes]{kuthesis}
\usepackage{graphicx}
\usepackage{amsmath}
\usepackage[utf8x]{inputenc}
\usepackage{subfig}
\usepackage[section]{placeins}
\graphicspath{ {images/} }
\usepackage{xspace}
\usepackage{epstopdf}

\newcommand \COMMENT  [1] {}       

\begin{document}

\Title{DEEP LEARNED FRAME PREDICTION \\ FOR VIDEO COMPRESSION} 
\Author{Serkan Sülün} 
\Year{November 22, 2018}
\Program{Electrical and Electronics Engineering}
\Signature{Prof. A. Murat Tekalp}
\Signature{Assoc. Prof. Engin Erzin}
\Signature{Prof. Bilge Günsel}

\prelimpages
\titlepage
\thesissignaturepage
\dedication{To my family and friends}
\abstract{
Motion compensation is one of the most essential methods for any video compression algorithm. Video frame prediction is a task analogous to motion compensation. In recent years, the task of frame prediction is undertaken by deep neural networks (DNNs). In this thesis we create a DNN to perform learned frame prediction and additionally implement a codec that contains our DNN. We train our network using two methods for two different goals. Firstly we train our network based on mean square error (MSE) only, aiming to obtain highest PSNR values at frame prediction and video compression. Secondly we use adversarial training to produce visually more realistic frame predictions. For frame prediction, we compare our method with the baseline methods of frame difference and 16x16 block motion compensation. For video compression we further include x264 video codec in the comparison. We show that in frame prediction, adversarial training produces frames that look sharper and more realistic, compared MSE based training, but in video compression it consistently performs worse. This proves that even though adversarial training is useful for generating video frames that are more pleasing to the human eye, they should not be employed for video compression. Moreover, our network trained with MSE produces accurate frame predictions, and in quantitative results, for both tasks, it produces comparable results in all videos and outperforms other methods on average. More specifically, learned frame prediction outperforms other methods in terms of rate-distortion performance in case of high motion video, while the rate-distortion performance of our method is competitive with that of x264 in low motion video.
}

\oz{
Tüm video sıkıştırma algoritmaları içinde en önemli metotlardan biri haraket dengelemedir. Video çerçevesi tahmini, hareket dengeleme ile benzer bir problemdir. Son yıllarda çerçeve tahmini, derin sinir ağları (DSA) tarafından yapılmaktadır. Bu tezde öğrenilmiş çerçeve tahmini yapmak için bir DSA yaratıyoruz ve ek olarak bu DSA'yı kapsayan bir video kodlayıcı üretiyoruz. DSA'mızı iki farklı amaç doğrultusunda iki farklı yöntem ile eğitiyoruz. İlk olarak çerçeve tahmini ve video sıkıştırmada en yüksek doruk sinyal gürültü oranı (PSNR) değerlerini elde etmek amacıyla DSA'mızı ortalama karesel hatayı (OKH) baz alarak eğitiyoruz. Ardından görsel olarak daha gerçekçi çerçeve tahminleri yapmak için çekişmeli eğitim yöntemini kullanıyoruz. Çerçeve tahmininde, yöntemimizi, referans yöntemler olan çerçeve farkı ve 16x16 blok hareket dengeleme ile karşılaştırıyoruz. Video sıkıştırmada, karşılaştırmaya x264 video kodlayıcıyı da dahil ediyoruz. Çerçeve tahmininde, çekişmeli eğitimin, OKH ile eğilmiş DSA'ya göre daha keskin ve gerçekçi çerçeveler ürettiğini; ancak video sıkıştırmada sürekli olarak daha başarısız olduğunu gösteriyoruz. Bu sonuç, çekişmeli eğitimin insan gözüne daha hoş görünen video çerçeveleri üretmesine rağmen video sıkıştırmada kullanılmaması gerektiğini kanıtlıyor. Ek olarak, OKH ile eğitilen DSA, çerçeve tahmininde yüksek doğruluklu sonuçlar üretiyor; nicel sonuçlarda, iki problem için de, tüm videolarda diğer yöntemlerle kıyaslanabilir sonuçlar veriyor ve ortalama başarıda diğer yöntemleri geçiyor. Daha detaylı olarak, yüksek hareketli videolarda, öğrenilmiş çerçeve tahmini, bithızı-bozulma performansında diğer yöntemleri geçiyor; ve düşük hareketli videolarda x264 ile yarışabilir bir sonuç üretiyor.
}
\acknowledgments{Firstly I would like to thank my supervisor Prof. A. Murat Tekalp for his academic and technical guidance, positivity, patience and ceaseless laughter. I also thank my office mates Selin, Tuğtekin, Tolga, Akın and Berker for their intellectual contributions and especially Arda, Ogün and Gonca for their deep friendship and drinking ability. In addition I want to thank my friends far away, Doğan, Felister and Melek for their never ending support. To my family Lale, Ali and Zeynep, I can not thank enough, for their unconditional love and compassion, for granting me freedom to follow any path I want, for spending almost all of their time with me as a child, and for giving me a happy and peaceful home, amongst plants and animals, in a town by the mountains and the sea. Finally I thank my dearest Ceren for her unbounded love.}

\tableofcontents
\listoftables
\listoffigures
\abbreviations{
DNN: Deep neural network

ANN: Artificial neural network

CNN: Convolutional neural network

GAN: Generative adversarial network

MSE: Mean square error

PSNR: Peak signal to noise ratio

GPU: Graphical processing unit

CPU: Central processing unit

SR: Super-resolution

SISR: Single-image super-resolution

EDSR: Enhanced deep super-resolution network

LSTM: Long short-term memory

SSIM: Structural similarity index measure

VAN: Visual analogy network

FP: Frame prediction

LFP: Learned frame prediction

ReLU: Rectified linear units

Tanh: Hyperbolic tangent

BPG: Better portable graphics

MPEG: Moving picture experts group

VBR: Variable bitrate

CBR: Constant bitrate

QP: Quantization parameter

BD: Bjontegaard delta

FD: Frame difference

MC: Motion compensation

NAL: Network abstraction layer

HRD: Hypothetical reference decoder
}
\textpages
%


\chapter{Introduction}

The transition from analog to digital media and the widespread adoption of the Internet as the premiere channel for media communication was made possible by advances in digital video processing. Sending digital video through a channel with limited bandwidth is a fundamental problem faced in video streaming over the Internet. A raw, high definition (1080p) video of 1 second occupies around 370 megabytes, while the highest national average of internet bandwidth is less than 8 megabytes per second. As a solution to this problem, \textit{video compression} deals with reducing the file size while maintaining visual integrity.

The most essential technique used in video compression is \textit{motion compensation}, which is done by dividing each video frame in blocks, estimating how these blocks move between frames and finally using this information to reconstruct each frame. Naturally this reconstruction is imperfect, so the \textit{residual error} needs to be transmitted alongside the \textit{motion vectors}. These two components constitute a large portion of the compressed video file. The sizes of these components can be reduced at the cost of less accurate reconstruction, creating videos with worse visual quality.
 
Until recently, the task of motion compensation, like all other signal processing problems, has been tackled using handcrafted methods. But in the last 8 years the human-engineered signal processing methods are being outperformed. With the advent of the modern graphical processing units (GPUs), artificial intelligence is more popular then ever. Using the computational power of these new GPUs, it only takes hours to create models that would require months ten years ago. These models of artificial neural networks (ANNs) can have hundreds of times more parameters than their older versions, including many layers of neurons and so we call them deep neural networks (DNNs). The usage of DNNs has prompted a subfield of machine learning, named \textit{deep learning}.

Having a vast number of neurons has greatly increased the computational and functional capacities of the ANNs. Comparing to human-engineered methods, DNNs are much more effective in modeling the real world phenomena. Almost every day, we use a DNN as a solution of a new engineering problem. Furthermore, techniques using DNNs are becoming the state-of-the-art in many fields of engineering.

A special type of ANN called \textit{convolutional neural network} (CNN) which models the human visual system, became a standard tool in computer vision, image and video processing tasks. Using these networks to create novel samples of data such as sounds and images, gave rise to another category of ANNs, named \textit{generative networks}. These networks can be used in a common task in video processing and computer vision, named \textit{video frame prediction}, namely generating future frame(s) of a video sequence, given its past frame(s). 

This problem is never deterministic. Consider a video of a person waving her hand. It is not possible to know if she will raise her hand by 30, 40 or 50 centimeters. In fact, any movement that she will make can be represented as a probability distribution. If our goal is to maximize our prediction's accuracy, the best prediction would be the expected value (average) of this distribution. In the context of image generation, producing an averaged image yields a blurry output.

In order to create sharp images that look pleasing to the human eye, the goal must be different than maximizing accuracy measured by the mean square error (MSE). Moreover there is no handcrafted metric that perfectly reflects human evaluation. \textit{Generative adversarial networks} (GANs) tackle this issue by using a second ANN, named \textit{discriminator}, as a substitute for the human critic. The first ANN, named \textit{generator} is trained to maximize scores given by the discriminator. Many of the works on video frame prediction use stochastic methods such as GANs, to create realistic frames, at the cost of reduced prediction accuracy.

Motion is almost always present in video sequences, hence this problem of video frame prediction is analogous to motion compensation. However, in frame prediction we do not employ the original (ground-truth) frame to be compressed even at the encoder. Learned frame prediction computes a predicted frame by only using the previous frames. Our goal is to perform predictive video compression by replacing the traditional motion compensation algorithm with a generative DNN for learned frame prediction. This method makes motion vectors redundant since a single DNN can perform frame prediction for any video. Consequently, compressed video occupies less space, or this extra space can be used to increase video quality.

We create and train DNNs to perform frame prediction on natural video sequences, with the highest possible accuracy. We also train a GAN to obtain sharper and more realistic images. Moreover, we build a codec encapsulating these DNNs, for the task of video compression and present a comparative analysis including the state-of-the-art video compression methods.


\chapter{Background and Related Work}


\section{Background on Neural Networks and Deep Learning}

Artificial neural networks (ANNs) are self-learning computational systems inspired by the animal brain. We use these networks to build approximations of unknown functions that take \textit{inputs} and create \textit{outputs}. 

\begin{figure}[h]
	\centering
	\includegraphics[scale=0.7]{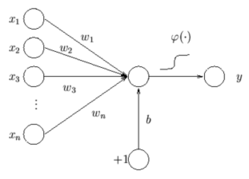}
	\caption{Perceptron}
	\label{fig:neuron}
\end{figure}

In the simplest case of a single neuron, \textit{perceptron}, multiplies all inputs $\{x_1, x_2, ..., x_n\} $ with its \textit{weights} $\{w_1, w_2, ..., w_n\} $, and then sums them up. Next, a \textit{bias} term $b$ is added and the result is then put through a non-linear function $f(.)$ to obtain the final output $y$, as displayed in Figure \ref{fig:neuron} and formulized in Equation~\ref{eqn:neuron}. 
\begin{equation}
\label{eqn:neuron}
y = f\bigg(\sum_{i=1}^{n} w_i x_i + b\bigg)
\end{equation}

\begin{figure}[h]
	\centering
	\includegraphics[scale=0.5]{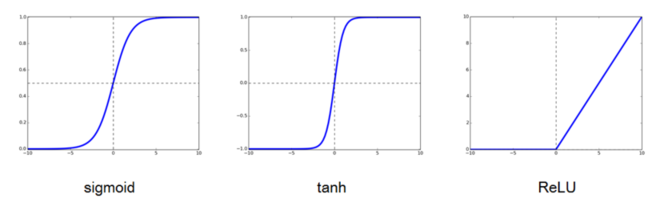}
	\caption{Non-linear functions}
	\label{fig:nonlinear}
\end{figure}

Since multiplying by a weight and adding a bias are linear operations, including a non-linear function is essential for better approximations. Figure \ref{fig:nonlinear} shows some popular examples of non-linear functions.

\begin{figure}[h]
	\centering
	\includegraphics[scale=0.3]{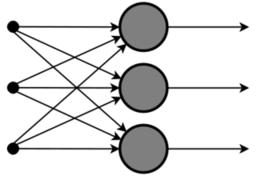}
	\caption{A layer with 3 inputs and 3 outputs}
	\label{fig:single_layer}
\end{figure}

In the case of multiple outputs, neurons can be stacked together, to create a \textit{layer}, as shown in Figure \ref{fig:single_layer} and formulized in Equation~\ref{eqn:layer} and \ref{eqn:layermatrix}.
\begin{equation} 
\label{eqn:layer}
y_j = f\bigg(\sum_{i=1}^{n} w_{ij} x_i + b_j\bigg) \qquad j = \{0,1,...,n\}
\end{equation}
Or in matrix notation:
\begin{equation}
\label{eqn:layermatrix} 
\begin{aligned}
y &= f(W x + b) \\
y &= \begin{bmatrix}
y_1 & y_2 & ... & y_m
\end{bmatrix}^T \\
b &= \begin{bmatrix}
b_1 & b_2 & ... & b_m
\end{bmatrix}^T \\
x &= \begin{bmatrix}
x_1 & x_2 & ... & x_n
\end{bmatrix}^T \\
W &= \begin{bmatrix}
w_{11} & w_{12} & \dots & w_{1n} \\
w_{21} & w_{22} & \dots & w_{2n} \\
\vdots & \vdots & \ddots & \vdots \\
w_{m1} & w_{m2} & \dots & w_{mn} 
\end{bmatrix}
\end{aligned}
\end{equation}

We call these layers of neurons a \textit{fully connected layer} since every input affects every output. In a single layer with $n$ inputs and $m$ outputs, there are $m \times n$ weights and $m$ biases.

\begin{figure}[h]
	\centering
	\includegraphics[scale=0.4]{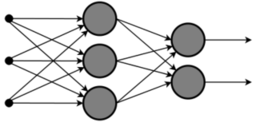}
	\caption{A two layer network with 3 inputs 2 outputs}
	\label{fig:two_layers}
\end{figure}

To approximate higher order functions, we stack these layers together and create a \textit{neural network}. Figure \ref{fig:two_layers} shows an example.

This operation can be written as: 
\begin{equation}
y = g(W_2(f(W_1+b_1))+b_2)
\end{equation}
where $g$ is another non-linear function.

\begin{figure}[h]
	\centering
	\includegraphics[scale=0.9]{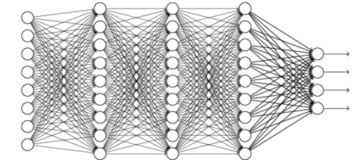}
	\caption{A deeper network with 8 inputs, 4 outputs and 4 layers.}
	\label{fig:dnn}
\end{figure}

To create more complex models, we need to stack more layers, as in Figure \ref{fig:dnn}. The number of layers indicate how \textit{deep} a neural network is.

\subsection{Convolutional Neural Networks}

Convolutional neural networks (CNN) \cite{cnn} are a class of ANNs inspired by the animal visual cortex and mostly used in visual data analysis. In CNNs, weights are grouped into \textit{kernels} (or \textit{filters}). These kernels are \textit{convolved} with the input data. Video and colored image data are 3 dimensional, having \textit{width}, \textit{height} and \textit{channels}. 
A single convolutional layer which takes a 3 dimensional input has 4 dimensional weights, with \textit{input channels}, \textit{output channels}, \textit{kernel width} and \textit{kernel height}. We define convolution as
\begin{equation}
y(x,y,n) = \sum_{c=0}^{C} \sum_{p=0}^{k} \sum_{q=0}^{k} I(x+p,y+q,c) \cdot w(c,n,p,q)
\end{equation}
Here $y(x,y,n)$ denotes the pixel value of the output image $y$'s $n$th channel, at coordinates $(x,y)$,  and $k$ denotes the side length of a square kernel. Just as fully connected layers, we stack convolutional layers to create deeper networks.

\subsection{Generative Networks}

Networks with a fully connected layer at its output layer produce a vector. In classification of a colored image, the output is a one dimensional vector of probabilities for each class, while the input is three dimensional. For each input sample, the goal is to find a suitable class \textit{label}. We train the network based on labeled data, which is called \textit{supervised learning}.

In contrast, generative networks specialize in creating outputs similar to its inputs. Since the goal is to produce data rather than labels, we train them based on unlabeled data, which is called \textit{unsupervised learning}.

\subsubsection{Autoencoders}

\begin{figure}[h]
	\centering
	\includegraphics[scale=0.6]{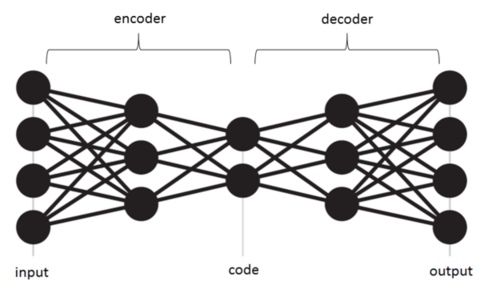}
	\caption{An autoencoder}
	\label{fig:ae}
\end{figure}

Autoencoders are generative networks aiming to create an output that is the same as the input. This is usually done to create a compact representation of the data. Autoencoders consist of two modules named \textit{encoder} and \textit{decoder}. The encoder takes the input and creates an output which is smaller in size. The output of the encoder is called \textit{code}. The decoder takes the code and creates an output with the same size as the input. Figure \ref{fig:ae} shows an example.

\subsubsection{Variational Autoencoders}

\begin{figure}[h]
	\centering
	\includegraphics[scale=0.4]{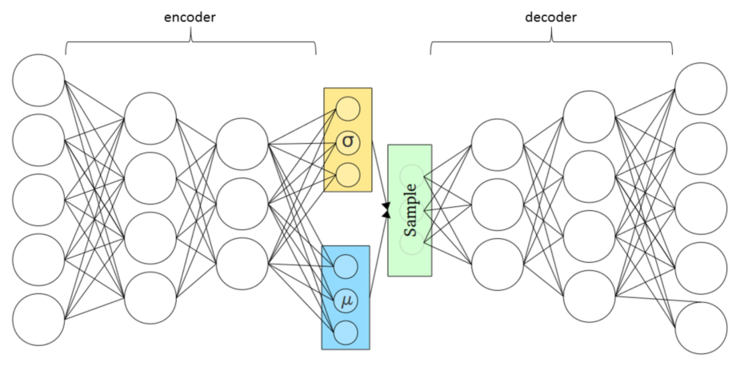}
	\caption{A variational autoencoder}
	\label{fig:vae}
\end{figure}

Variational Autoencoders \cite{vae} are capable of creating novel outputs that are similar to the input data. They achieve this by modeling the code as a Gaussian random variable. Figure~\ref{fig:vae} shows an exemplary model. Half of the encoder's output constructs the mean and the other half constructs the standard variation of the code. Using these two parameters, we sample a random variable and feed it to the decoder. During deployment, we don't use the encoder, instead we simply obtain the sample from the Gaussian distribution and feed it into the decoder. 

\subsubsection{Generative Adversarial Networks}

Generative adversarial networks (GANs) \cite{gan} are able to create authentic samples of data. This is achieved by using two competing networks, the \textit{generator} and the \textit{discriminator}. 

The discriminator's task is to differentiate between the original data (real) and data generated by the generator (fake). Hence its output is a single number between 0 and 1, indicating the probability that its input is real. During training, we use both real and fake samples as inputs, with their target labels being 1 and 0 respectively.

The generator's task is to create outputs that will trick the discriminator. It is trained based on the discriminator's feedback, where the target value is 1 (real). To make sure that it produces different outputs, we use random noise as its input.

\begin{figure}[h]
	\centering
	\includegraphics[scale=0.9]{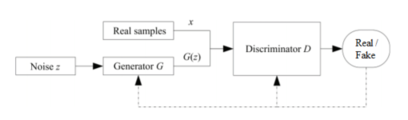}
	\caption{Generative Adversarial Network's block diagram}
	\label{fig:gan}
\end{figure}

A block diagram can be seen in Figure \ref{fig:gan}.

\section{Related Work on Video Frame Prediction}
\label{related}
\begin{figure}[h]
	\centering
	\includegraphics[scale=0.4]{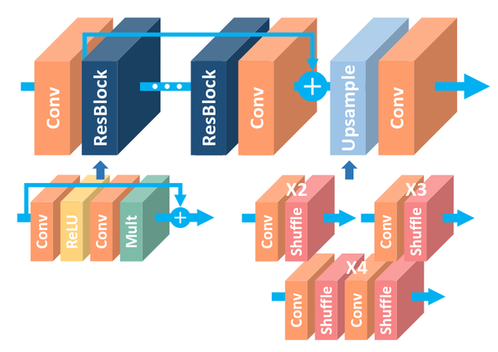}
	\caption{EDSR model}
	\label{fig:edsr}
\end{figure}

We first present a paper written for the task of image super-resolution and not for video frame prediction, since our model has a very similar structure as explained in Section~\ref{generator}. Lim et al. have created a deep convolutional residual network named Enhanced Deep Super-Resolution Network (EDSR) \cite{edsr}, winning the NTIRE 2017 Challenge on Single Image Super-Resolution \cite{ntire}. Their network takes a single colored low resolution image as input and reproduces this image at a higher resolution. To minimize error, they use a deep residual convolutional network. All intermediate layer's activation sizes (height and width) are the same as the input's, to prevent a loss of spatial information. In order to create the output image at a higher resolution they use an upsampling block with a pixel shuffling layer \cite{pixelshuffler}. Their model can be seen in Figure \ref{fig:edsr}.

Next, works on video frame prediction are presented chronologically and finally they are classified under various categories. We can classify these models based on their training losses and methods. The most commonly used losses are mean square error, $\ell^1$ distance and cross-entropy loss, but some works define a novel loss function e.g by taking a difference of image gradients. Stochastic methods use variational encoders or GANs, employing KL-divergence or adversarial loss, respectively. Conceptual loss is also used, by comparing frames at the feature space, using a pretrained network. Another way to categorize these networks is by their output domain, namely pixel level, patch level and frame level generation. Finally, we can categorize them by their method of generation, frame reconstruction and frame transformation. We note that many of these works fit into multiple categories. We also encourage interested readers to examine two surveys \cite{survey, survey_ge} written on video frame prediction using DNNs.

\begin{figure}[h]
	\centering
	\includegraphics[scale=0.45]{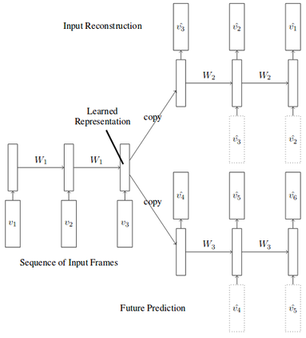}
	\caption{Srivastava's model}
	\label{fig:srivastava}
\end{figure}

Srivastava et al. \cite{srivastava} work on two tasks, input frame reconstruction and future frame generation using long short-term memory (LSTM) autoencoders. They also come up with a composite model, performing reconstruction and prediction at the same time, outperforming the model that only predicts the future frames. They further improve their results by conditioning the decoder using the output from previous time-step. Their best performing model can be seen in Figure \ref{fig:srivastava}.

\begin{figure}[h]
	\centering
	\includegraphics[scale=0.3]{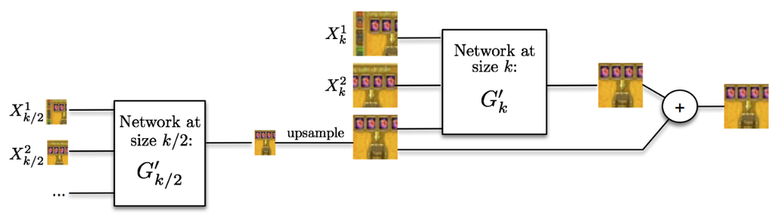}
	\caption{Mathieu's multiscale architecture}
	\label{fig:beyondmse}
\end{figure}

Mathieu et al. \cite{mathieu} aim at generating sharper images. Their first contribution is to use a multi-scale architecture, concatenating the output of the smaller scale's network with the input of the larger scale's network, as seen in Figure \ref{fig:beyondmse}. Their second contribution is to use different loss functions, complementary to the mean square loss. They introduce gradient difference loss by calculating the difference between the gradients of predicted and ground-truth images. They also use a multi-scale discriminator network to incorporate adversarial loss. The training loss is a weighted sum of mean square, gradient difference and adversarial losses.

\begin{figure}[h]
	\centering
	\includegraphics[scale=0.7]{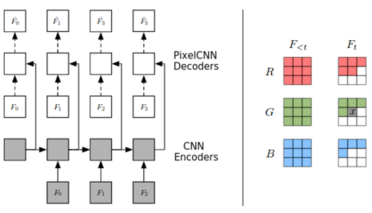}
	\caption{Video Pixel Networks}
	\label{fig:vpn}
\end{figure}

Kalchbrenner et al. \cite{kalchbrenner} model the pixel values in four dimensions (time, height, width, color) directly and causally. Their model, named Video Pixel Networks, has an encoder-decoder structure. The encoder part consists of CNNs with dilated convolutions, followed by a convolutional LSTM. This part of the network models temporal dependencies, and its output conditions the decoder. The decoder consists of a PixelCNN \cite{pixelcnn} which models individual pixels sequentially. Using masked convolutions, PixelCNN can use previously generated pixels within the same image and even the pixel at the same location from the previous color channel, as inputs. Implicitly, the decoder can access the previous frames since the output of the encoder is passed to the decoder. Their model can be seen in Figure \ref{fig:vpn}.

\begin{figure}[h]
	\centering
	\includegraphics[scale=0.4]{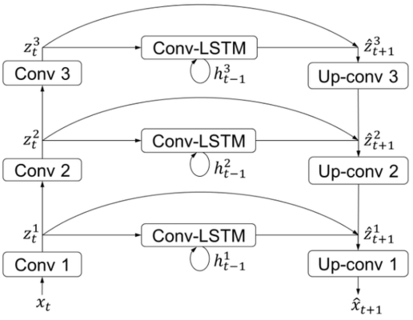}
	\caption{Video Ladder Networks}
	\label{fig:vln}
\end{figure}

Cricri et al., in their work named Video Ladder Networks \cite{cricri} use a multi-scale encoder-decoder architecture. Encoder and decoder consist of cascaded upscaling and downscaling convolutional layers respectively. Encoder's activations at each scale are transmitted to the decoder's convolutional layer at the corresponding scale, through two lateral connections, a recurrent, and a feedforward (skip) connection. The recurrent connections employ convolutional LSTMs, capable of using information from the past frames. The feedforward connection only sends features from the current frame, hence it is useful for the modeling of the static image parts. Their baseline architecture can be seen in Figure \ref{fig:vln}. They extend this model by replacing the single convolutional layers with residual convolutional blocks.

\begin{figure}[h]
	\centering
	\includegraphics[scale=0.4]{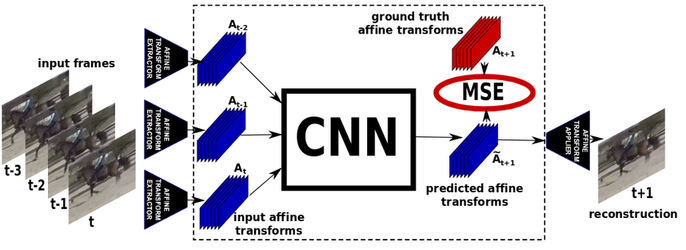}
	\caption{Amersfoort's model}
	\label{fig:amersfoort}
\end{figure}

Instead of modeling pixels, Amersfoort et al. \cite{amersfoort} build a CNN that only predicts the affine transformations between the patches from consecutive frames. The network takes the affine transformation matrices from the previous frames as input and outputs a predicted affine transformation matrix for the next frame. This network is trained using mean square loss between the predicted and ground-truth affine transformation matrices. The predicted affine transformation is later applied on the current frame to generate the next one. Their model can be seen in Figure \ref{fig:amersfoort}.

\begin{figure}[h]
	\centering
	\includegraphics[scale=0.3]{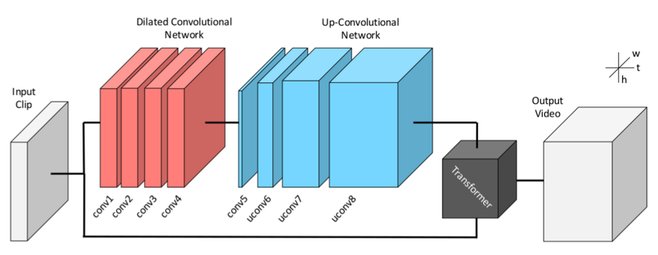}
	\caption{Vondrick's model}
	\label{fig:vondrick}
\end{figure}

Vondrick et al. \cite{vondrick} also aim at predicting frame transformations but train the entire model end-to-end. A CNN takes the input frames and outputs a transformation matrix. The transformation matrix is used to warp the last frame. Using this matrix, for each pixel,  a weighted average of its neighboring pixels is computed. This is implemented as the standard convolution operation. The model is seen in Figure \ref{fig:vondrick}. To obtain sharp predictions, the network is trained using adversarial loss only. They have measured the model's performance by using it as a pretraining method for object classification tasks, and with subjective evaluation. No peak signal-to-noise ratio (PSNR) or structural similarity index measure (SSIM) results are reported.

\begin{figure}[h]
	\centering
	\includegraphics[scale=0.4]{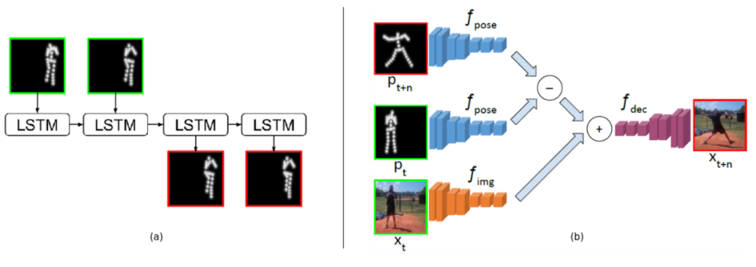}
	\caption{Villegas' model. (a) Pose predicting LSTM (b) Frame predicting CNN}
	\label{fig:villegas}
\end{figure}

Villegas et al. \cite{villegaslongterm} make long term predictions on human videos, while avoiding the frame by frame accumulation of error, namely drift. With the help of a pretrained network \cite{pose}, they extract human pose information from each frame. Using the pose information as input, an LSTM autoencoder is trained, specifically on future pose prediction. The last frame, its corresponding pose and the future pose are fed into a visual analogy network (VAN) \cite{van}, that is a CNN with arithmetic operations, to generate the future frame. The loss function for this network consists of the sum of $\ell^2$, perceptual and adversarial losses. Perceptual loss is obtained by measuring the distance between the generated and ground-truth images, in a feature space created by a pretrained classification network \cite{perceptualloss}. The only recurrent predictions are done by the pose predicting network, hence a predicted pose at an arbitrary future time-step can be used to generate the corresponding frame, without generating its earlier frames. The model can be seen in Figure \ref{fig:villegas}.

\begin{figure}[h]
	\centering
	\includegraphics[scale=0.4]{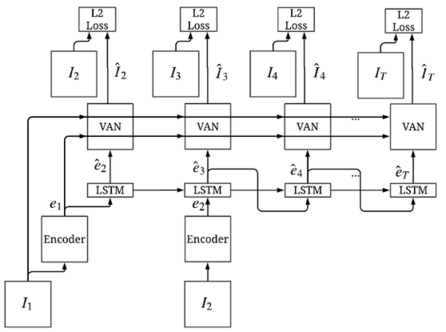}
	\caption{Wichers' end-to-end model}
	\label{fig:wichers}
\end{figure}

Following up the work of Villegas, Wichers et al. \cite{wichers2018hierarchical} replace the pre-trained pose extracting network with a trainable encoder, enabling end-to-end training. An LSTM is used to predict the encoding of the next frame, given the encoding of the current frame. The model can be seen in Figure \ref{fig:wichers}. They have performed three experiments. Firstly, they trained the entire network end-to-end. With this method, they have observed that VAN produces blurry images when the predictor LSTM is not confident. To address this issue they have trained the predictor and VAN separately, obtaining sharper outputs. To achieve even sharper results, they have included adversarial loss during the training of the predictor. It should be noted that a direct quantitative evaluation of generated images using standard metrics such as PSNR or SSIM is not presented.

\begin{figure}[h]
	\centering
	\includegraphics[scale=0.3]{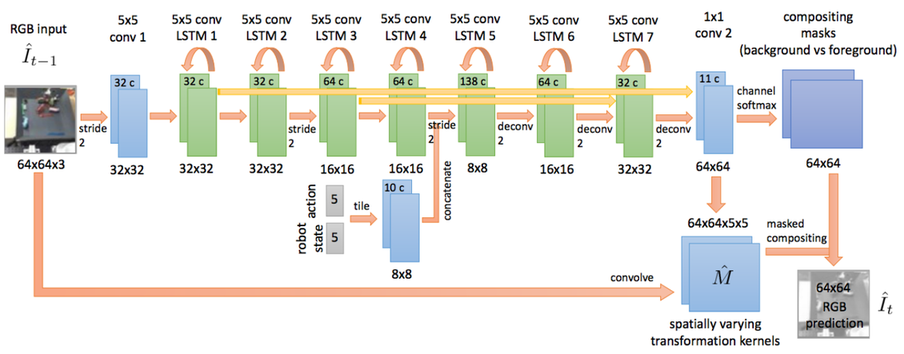}
	\caption{Finn's model}
	\label{fig:finn}
\end{figure}

Finn et al. \cite{finn} focus on predicting object motions and then transforming the video frames rather than reconstructing them. In order to model motion explicitly, they use convolutional LSTMs. Their models are used on Human3.6M and robotic pushing datasets. For robotic pushing dataset, robot states and actions are also used as inputs by concatenating with the activation of the middle layer. The size of the activation is compatible with the state-action input, which is smaller than the rest of the activations. This is achieved by using an encoder-decoder structure, with decreasing activation sizes until the middle of the network and increasing activation sizes afterwards. Their models create motion transformation kernels to apply motion transformation to the input image, and they create compositing masks to combine multiple predictions into a single image. Their model, can be seen in Figure \ref{fig:finn}.

\begin{figure}[h]
	\centering
	\includegraphics[scale=0.35]{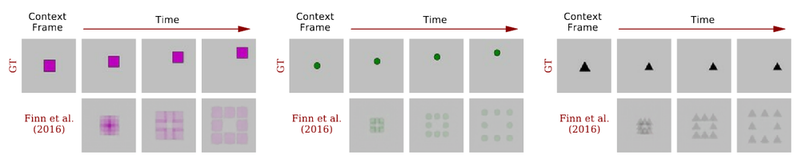}
	\caption{Qualitative results using a single frame as input to Finn's model}
	\label{fig:finn_blur}
\end{figure}

\begin{figure}[h]
	\centering
	\includegraphics[scale=0.3]{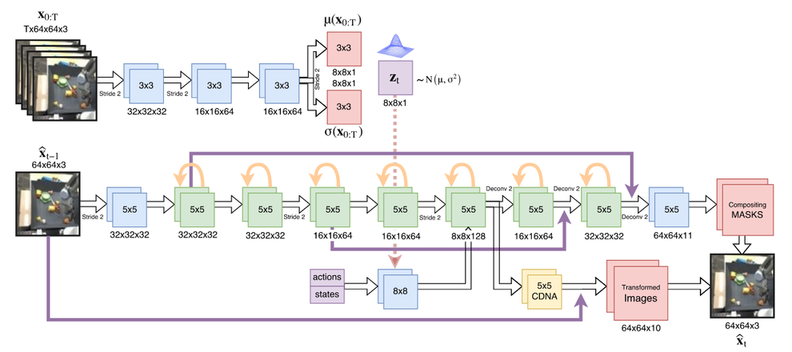}
	\caption{Babaeizadeh's model}
	\label{fig:baba}
\end{figure}

In a follow-up work, Babaeizadeh et al. \cite{babaeizadeh2017stochastic} show that, given a single input frame, Finn's model produces a blurry output, including artifacts from all possible future outcomes, as show in Figure~\ref{fig:finn_blur}. To avoid this, they supplement Finn's model with a variational encoder. The output of the encoder is used to sample a latent variable which is fed into the generator. This sampling procedure is equivalent to choosing a single mode from the multimodal distribution representing possible future outcomes. The model is shown in Figure \ref{fig:baba}.

\begin{figure}[h]
	\centering
	\includegraphics[scale=0.6]{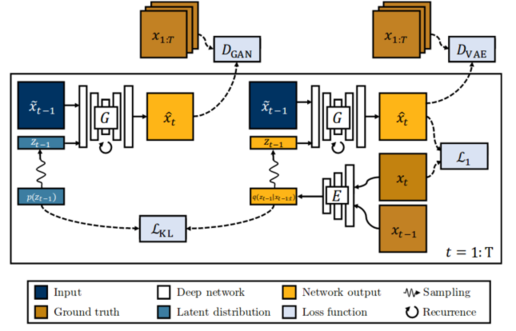}
	\caption{Lee's model}
	\label{fig:lee}
\end{figure}

In the next follow-up work, Lee et al. \cite{lee2018stochastic} supplement Babaeizadeh's model with an adversarial loss, to obtain more realistic results. The model can be seen in Figure \ref{fig:lee}.

Denton et al. \cite{denton2018stochastic} use a variational encoder to deal with uncertainty. The output of the encoder is a probability distribution of the potential outcomes for the predicted frame. A latent variable is sampled from this distribution, and together with the input frame, they are fed into the decoder. They use convolutional LSTMs to enable training at each time-step for a variable number of frames.

\begin{figure}[h]
	\centering
	\includegraphics[scale=0.3]{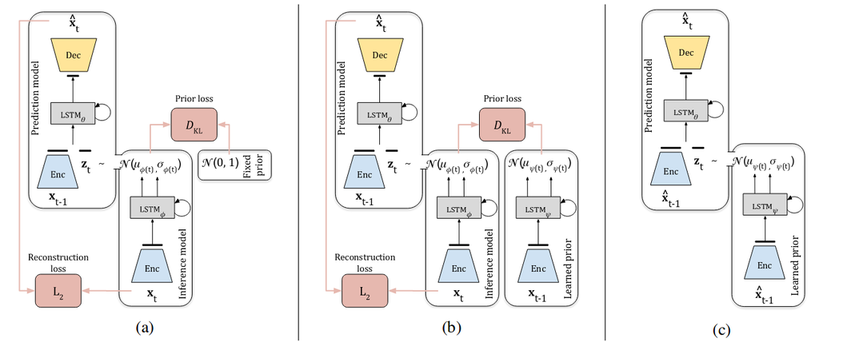}
	\caption{Denton's model. (a) Training with vanilla VAE (b) Training with conditional VAE (c) Testing with conditional VAE}
	\label{fig:denton}
\end{figure}

They present two models, using a vanilla variational encoder and a conditional variational encoder. The former models the uncertainty as a zero mean unit variance Gaussian distribution while the latter creates a Gaussian distribution conditioned on the previous frames. Their models at training and test times can be seen in Figure \ref{fig:denton}.

\clearpage

\subsection{Categorization}

In this section, related works are classified under categories defined in the beginning of Section \ref{related}.

\textbf{Loss/training method}

\textit{Mean square error:} Srivastava \cite{srivastava}, Mathieu \cite{mathieu}, Amersfoort \cite{amersfoort}, Villegas \cite{villegaslongterm}, Wichers \cite{wichers2018hierarchical}, Finn \cite{finn}, Babaeizadeh \cite{babaeizadeh2017stochastic}, Lee \cite{lee2018stochastic}, Denton \cite{denton2018stochastic} 

\textit{$\ell^1$ distance:} Mathieu \cite{mathieu}

\textit{Cross entropy:} Srivastava \cite{srivastava}, Kalchbrenner \cite{kalchbrenner}, Cricri \cite{cricri}

\textit{Gradient difference:} Mathieu \cite{mathieu}

\textit{Perceptual loss:}  Villegas \cite{villegaslongterm}

\textit{Adversarial loss/GAN:} Mathieu \cite{mathieu}, Vondrick \cite{vondrick}, Villegas \cite{villegaslongterm}, Lee \cite{lee2018stochastic} 

\textit{KL divergence/Variational:} Babaeizadeh \cite{babaeizadeh2017stochastic}, Lee \cite{lee2018stochastic}, Denton \cite{denton2018stochastic}

\textbf{Network output domain}

\textit{Frame level:} Mathieu \cite{mathieu}, Vondrick \cite{vondrick}, Villegas \cite{villegaslongterm}, Wichers \cite{wichers2018hierarchical}, Finn \cite{finn},  Babaeizadeh \cite{babaeizadeh2017stochastic}, Lee \cite{lee2018stochastic}, Denton \cite{denton2018stochastic} 

\textit{Patch level:} Srivastava \cite{srivastava}, Amersfoort \cite{amersfoort}

\textit{Pixel level:} Kalchbrenner \cite{kalchbrenner}

\textbf{Method}

\textit{Frame reconstruction:} Srivastava \cite{srivastava}, Mathieu \cite{mathieu}, Kalchbrenner \cite{kalchbrenner}, Cricri \cite{cricri}, Wichers \cite{wichers2018hierarchical}, Denton \cite{denton2018stochastic}

\textit{Frame transformation:} Amersfoort \cite{amersfoort}, Vondrick \cite{vondrick}, Villegas \cite{villegaslongterm}, Finn \cite{finn}, Babaeizadeh \cite{babaeizadeh2017stochastic}, Lee \cite{lee2018stochastic}


\chapter{Methodology}

The methodology proposed in this thesis for learned predictive video compression consists of two steps: learned video frame prediction, which is discussed in Section~\ref{framepred}, and compression of the predictive frame differences, which is described in Section~\ref{videocomp}.

\section{Learned Video Frame Prediction}
\label{framepred}
We first discuss the rationale for our choice of network architecture and training method for the purpose of frame prediction in predictive video compression in Section~\ref{architecture}. Next, we present the details of our generator and discriminator network architectures in Sections~\ref{generator} and~\ref{discriminator}, respectively. We end this section by explaining the details of training procedures using standard mean square loss only and combined mean square and adversarial loss in Section~\ref{training}.

\subsection{Choice of Architecture and Training Method} 
\label{architecture}
Video frame prediction is a stochastic problem, where the target frame can be modeled by a multimodal probability distribution, since there are more than one possible outcomes for the predicted frame. Standard methods aiming to minimize the mean square error converge to the expected value of this multimodal distribution, which is an average of all possible outcomes; hence, minimization of mean square error generally yields a blurry predicted image. 

Many approaches, which are reviewed in Section~\ref{related}, have been proposed in recent years to overcome this problem and generate sharper and more realistic looking predicted images. A majority of these works use recurrent models to process an arbitrary number of input and output frames. Until recently, LSTMs were the top choice of architecture to solve sequence learning problems, especially in speech and natural language processing, where recurrent models yield amazing results in speech to text conversion and machine translation. However, training LSTMs for video processing can still be problematic.

With the introduction of the ResNet architecture, which overcomes the vanishing gradients problem by clever bypass connections, it is easier to train deeper CNNs to learn temporal dynamics from a fixed number of past video frames. Feeding a fixed number of past frames to a CNN requires only modification of the input layer. Hence, the additional computational cost of processing a fixed-size block of input frames (instead of a single input frame) is insignificant  compared to the computational cost of the rest of the network. In theory, LSTMs can remember entire history of a video from the beginning. However, in practice, we obtain as good if not better performance (due to training issues) by batch processing of a fixed size block of past frames using a CNN rather than processing all frames sequentially using a recurrent architecture.

Another important factor that affects video compression performance is the choice of training methodology and loss function. It has been observed in the literature that stochastic sampling methods such as variational autoencoders, and adversarial training procedures similar to those used in training GANs do produce sharper predicted images with more spatial detail in frame prediction problems. However, they do so at the expense of higher mean square error when compared to the actual next frame. This raises the question: How do we evaluate the prediction performance of a neural network? In computer vision applications such as surveillance or robotics planning, where the predicted images are to be viewed or used as final product and/or there is no ground-truth available, then sharper and more natural looking images are preferred. In contrast, in video compression the predicted image is only an intermediate result (not to be viewed), and the goal is to minimize the bitrate required to send the difference between the predicted and actual frames (ground-truth). In this case, it turns out that the entropy of the predictive frame difference is lower when we train the network using the mean square loss even if the predicted images look blurry.

In conclusion, we chose to employ a deep convolutional network, where the activation sizes stay constant from the input to the output layer; hence, avoiding autoencoder (encoder-decoder) architectures, recurrent architectures, and stochastic sampling approaches. We train the CNN using the mean square loss function in order to obtain the minimum mean square prediction error for video compression task. In the following, we propose to employ a modified version of the Enhanced Deep Super-Resolution (EDSR) network \cite{edsr} as a generator for the task of video frame prediction. We only modified the input and output layers of the EDSR network as described below. Next, we propose a discriminator network for the purpose of adversarial training of the generator network. Finally, we discuss the details of training procedures used. 

\subsection{Generator Network} 
\label{generator}
Learned frame prediction (FP) problem is similar to the learned single-image super-resolution (SISR) problem in that they both generate new images using generative networks. In the SISR problem, the generative network predicts a high resolution image based on a learned model of high frequency content of images given a low resolution image, while in the FP problem, the generative network predicts the next frame based on a learned model of temporal dynamics given $N$ previous frames. 

The Enhanced Deep Super-Resolution (EDSR) network \cite{edsr} is the winner of the NTIRE 2017 Single Image Super-Resolution Challenge \cite{ntire}. Unlike many other works, which employ an autoencoder architecture for image generation, in the EDSR network the size of all intermediate layer activations is the same as that of the input layer. That is, they avert loss of spatial information by avoiding an encoder-decoder architecture. This property of the EDSR network architecture and its success in the SISR problem form the basis of our choice for generator architecture. 

The EDSR network takes a single color image and outputs a single color image with double the height and width. In contrast, our network's input layer takes $N$ past grayscale (Y-channel) frames and our output layer produces a single grayscale frame, whose size (height and width) is the same as the input frames. Therefore, we modified both the input and output layers of the EDSR network. Since the input and output images are all the same size, we don't need an upscaling layer that was used by the EDSR network. 

\begin{figure}[ht]
\centering
	\includegraphics[scale=0.46]{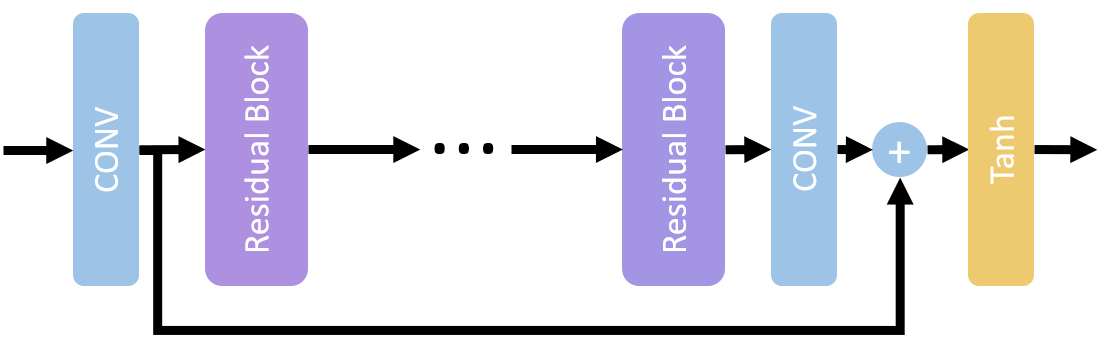} \vspace{-8pt} \\
	(a) \vspace{10pt }\\
	\includegraphics[scale=0.35]{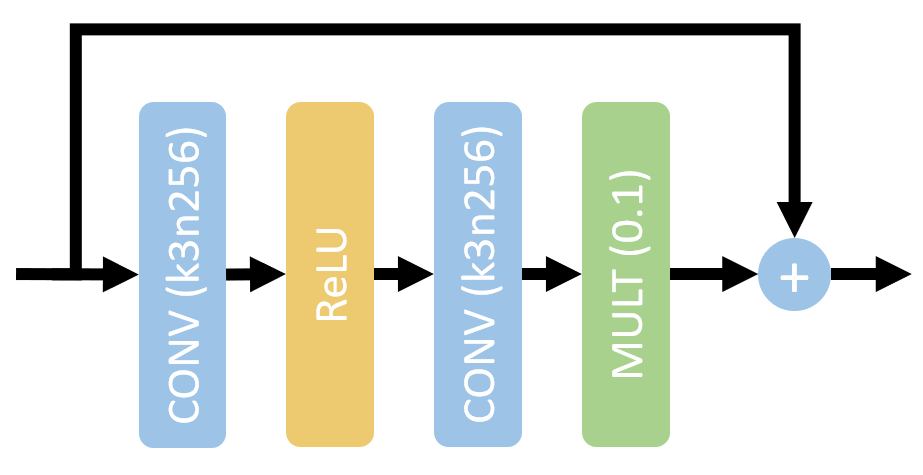} \vspace{-6pt} \\
	(b)
\caption{Block diagram of the generator network: (a) the EDSR network (b) each residual block of the EDSR network shown in purple in (a).}
\label{fig:resnet}
\end{figure}

The architecture of the generator network is depicted in Figure \ref{fig:resnet}. Our network has 32 residual blocks, each having 2 convolutional layers with kernel size 3, padding 1 and channel depth 256 and rectified linear units (ReLU) as the activation functions between hidden layers. The model is fully convolutional so it can process inputs with an arbitrary size (height and width). The height and width of all intermediate layer activations and the output layer are the same as those of the input layer. At the end of each residual block, residual scaling with 0.1 is applied \cite{residualscaling}. At network's output hyperbolic tangent (Tanh) is used as the nonlinear function, ensuring output values between -1 and 1. At test time, the outputs are scaled between 0 and 255 and converted to 8-bit unsigned integers to generate predicted video frames. 

We observe that the best results with EDSR in the NTIRE 2017 competition were obtained by using their largest network. Since our ultimate goal is to obtain the best rate-distortion performance in video compression, we aim at obtaining the lowest mean square prediction error. As will be demonstrated in Chapter~\ref{results}, the performance of the modified EDSR network to predict frames is surprisingly good. We observed in our FP experiments that using larger networks consistently provide lower mean square prediction error similar to that observed in SISR experiments. This is plausible since the task of image/video generation is highly complex so overfitting is rarely observed.

\subsection{Discriminator Network}
\label{discriminator}
A discriminator network is needed for adversarial training of the generator network in order to compare performances of training with MSE only and adversarial training, both in terms of visual quality of predicted frames as well as the resulting rate-distortion performance in predictive video compression.

Even though large and deep residual networks with a fully connected output layer have been used as the discriminator network in some adversarially trained super-resolution models, such as the SRGAN model \cite{srgan}, in our experiments we obtained better results when a smaller discriminator network is used. The discriminator undertakes a simpler task compared to the generator network, where the output and target of the discriminator network are single numbers, while the generator has a much more complicated task of producing full frames. As a result, using similar size networks typically results in the discriminator outperforming the generator, causing an instability in the process of training. To this effect, we used a small discriminator network with no fully connected layers. 

\begin{figure}[h]
	\centering
	\includegraphics[scale=0.4]{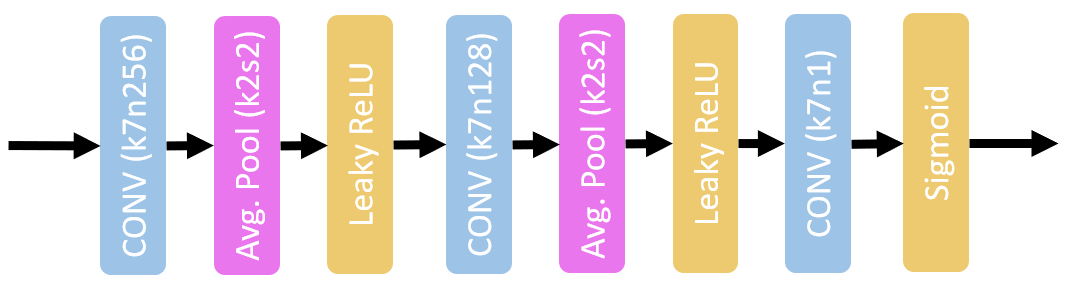}
	\caption{Block diagram of the discriminator network.}
	\label{fig:discriminator}
\end{figure}

The block diagram of our proposed discriminator network architecture is depicted in Figure \ref{fig:discriminator}. There are three convolutional layers with kernel size of 7 and varying channels depths. Even though the discriminator is fully convolutional, by using pooling layers with kernel size 2 and stride 2, and avoiding padding, we reduce the activation sizes throughout the network and obtain a single value in the output. We note that this is only possible when an input with size 48x48 is used, but this does not create any problems since the discriminator is only used during training. As recommended in \cite{ganhacks}, for a stable training, we have used average pooling instead of maximum pooling, to avoid sparse gradients. Between the hidden layers, leaky rectified linear unit (Leaky ReLU) with a slope of 0.2 is used as the nonlinearity, as advised in \cite{dcgan}. The final nonlinear function is a sigmoid, enabling an output value between 0 and 1, regarded as a probability of the input image's authenticity (higher means more realistic). 

\subsection{Training Details}
\label{training}
We first trained the generator network using the mean square loss. Then, we trained the discriminator network together with our pretrained generator network in order to perform adversarial training, where the discriminator is trained with adversarial loss only, while the generator is trained with a weighted combination of mean square and adversarial losses.

\subsubsection{Training Dataset}
We used UCF101 \cite{ucf101} dataset, created by the University of Central Florida, as our training set. This dataset contains 13,320 videos of human actions, collected from Youtube. We converted the videos into grayscale (Y-channel) and extracted sequences of patches with size $48 \times 48$ pixels and duration 9 frames. We predict the 9th frame given the first 8, and use the 9th frame of each patch sequence as the ground-truth.

While extracting patch sequences, we selected the video, the starting frame and patch location randomly, with the condition that there is motion between all consecutive pairs of patches within selected patch sequences. The pixel values range between 0 and 255, and we set the motion detection threshold as mean square difference of 7 between successive pairs of patches. With probability 0.05, we ignored this threshold and extracted the patch sequence anyway completely randomly. Overall, we extracted about 2 million patch sequences as our training dataset. We saved the extracted patch sequences as 8-bit unsigned integer to enable faster data transfer from the CPU to the GPU. At training time, we transfer the patches to the GPU, and then convert them to 32-bit floats by scaling the pixel values between -1 and 1.

With a training set of this size, we don't use the concept of training by \textit{epochs}, meaning that we don't make multiple passes over the dataset. Instead, we pick a set of patch sequences randomly from the entire patch sequence dataset to construct minibatches at training time, leading to the concept of training by \textit{steps} or \textit{iterations}, where one training step corresponds to processing one minibatch.

\subsubsection{Training with mean square loss}

We first trained our model based only on mean square loss. Since the model takes 8 frames as input and predicts the 9th frame, we compute the mean square error over the 9th patch, which is defined by
\begin{equation} \label{mse}
L_{MS} = \frac{1}{N} \sum_{i=1}^{N}(y_i - x_i)^2
\end{equation}
where $x$ and $y$ denote the generated and the ground-truth (original) 9th patch, respectively, and $i$ is the index looping over all $N$ pixels in the ground-truth patch.

We used Adam optimizer \cite{adam} with an initial learning rate of 1e-4 and a batch size of 32. If the training loss did not decrease for 6000 steps, we halved the learning rate. We trained our model for 400,000 steps (iterations), which lasted about 8 days using a single NVIDIA GeForce GTX 1080Ti GPU on a HP Server with Intel Xeon Gold CPU @ 2.30GHz with 24 cores. 

\subsubsection{Training with combined mean square and adversarial losses}

We next fine tune the parameters of the generator network by the process of adversarial training, which employs the discriminator network. To this effect, we jointly trained the randomly initialized discriminator and the pretrained generator networks. The minibatch sizes are 16 for the generator and 32 for the discriminator. A minibatch for training the discriminator network consists of 16 original (ground-truth) and 16 generated (fake) patch-sequence samples. 

Original samples are actual training patch sequences. Generated samples are composed of the first 8 frames of original patch sequences that are input to the generator network concatenated with the predicted 9th patch that is the output of the generator network. Thus, a generated sample includes the first 8 original patches and a generated image as the 9th patch. By feeding sequences instead of single frames into the discriminator, we enforce the discriminator to make its predictions based on sequence continuity. This procedure is similar to that was used by Mathieu et al. \cite{mathieu}. We use target labels of 0 and 1 for the generated and original samples, respectively. Then, the discriminator network is trained using the binary cross entropy loss, given by
\begin{equation}
L_{BCE} = -y\log x - (1-y)\log (1-x)
\end{equation}
where $x$ is the discriminator network's output (prediction) and $y$ is the binary ground-truth label.

The generator network is trained using a weighted combination of mean square and adversarial losses. While calculating the adversarial loss for the generator network, we only feed generated sequences to the discriminator and use a target label of 1, since the generator's aim is to create realistic images. Thus, the adversarial loss for the generator network is defined as
\begin{equation}
L^G_{ADV} = -\log x_{disc}
\end{equation}
Hence generator's combined loss becomes
\begin{equation}
L = \lambda_{MS} \left(\frac{1}{N} \sum_{i=1}^{N}(y_i - x_i)^2 \right) - \lambda_{ADV} \log x_{disc}
\end{equation}
where $x_i$ is a single pixel in generator's output, $x_{disc}$ is discriminator's output, $N$ is the number of pixels, $\lambda_{MS}$ and $\lambda_{ADV}$ are weights for the mean square and adversarial losses, respectively. We have used $\lambda_{MS} = 0.95$ and $\lambda_{ADV} = 0.05$ similar to that was done in \cite{mathieu}.

The learning rates are constant at 1e-6 and 1e-5 for the generator and discriminator, respectively. The adversarial training for 300000 steps took about 5 days.

\section{Compression of Predictive Frame Differences}
\label{videocomp}
Predictive coding is a common practice to exploit temporal redundancy in video compression. The standard method of exploiting temporal redundancy in the video compression community is motion compensation, where motion vectors computed at the encoder between variable sized blocks of the current (original) frame and a reconstructed reference frame are sent to the decoder as side information. While this techniques has been very successful for over three decades, it has some drawbacks:
1) Rate-distortion optimization to determine the best block partitioning and estimation of motion vectors requires significant computational resources at the encoder, which makes real-time encoding a challenge, and
2) sending motion vectors as side information incurs a bitrate overhead, which is significant in low bitrate video coding.

We propose to replace the motion compensation module in predictive video compression with learned frame prediction using a neural network. More specifically, we propose to encode the difference image between the ground-truth and predicted frames using a still image encoder. In this thesis, we employed the Better Portable Graphics (BPG) still image codec \cite{bpg} for this purpose.

\subsection{Encoder}
\begin{figure}[h]
	\centering
	\includegraphics[scale=0.55]{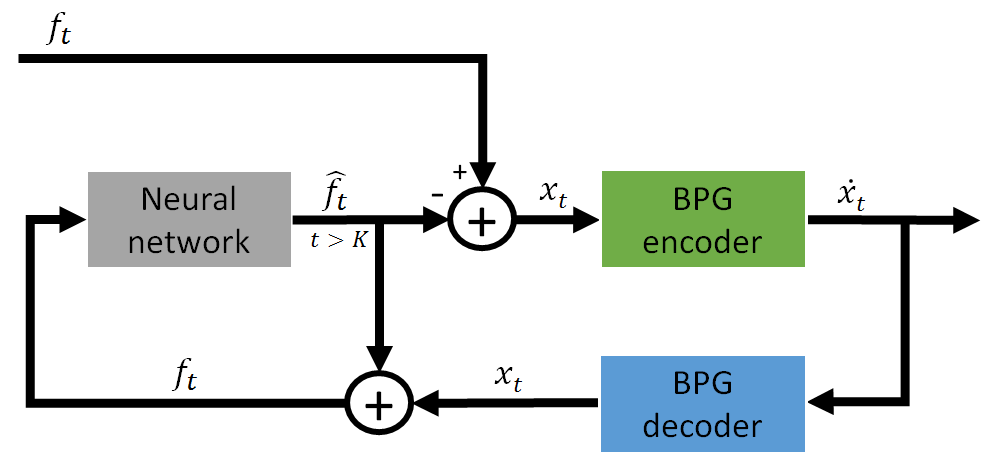}
	\caption{The block diagram of the proposed encoder.}
\label{fig:encoder}
\end{figure}

The block diagram of the proposed video encoder is depicted in Figure~\ref{fig:encoder}.
The first $K$ frames are input to the BPG encoder as I pictures without prediction. Starting with frame $K + 1$, the predicted frame is subtracted from the original frame, and the difference is encoded using the BPG encoder. In order to input exactly the same past frames into the neural networks in the encoder and decoder, the encoder has a BPG decoder in the feedback loop which produces the decoded frame difference. The decoded differences are added to the next frame predictions in order to produce decoded frames that are identical to the ones at the decoder, which become the input to the neural network for frame prediction at the next time step. 

\subsection{Decoder}
\begin{figure}[h]
	\centering
	\includegraphics[scale=0.55]{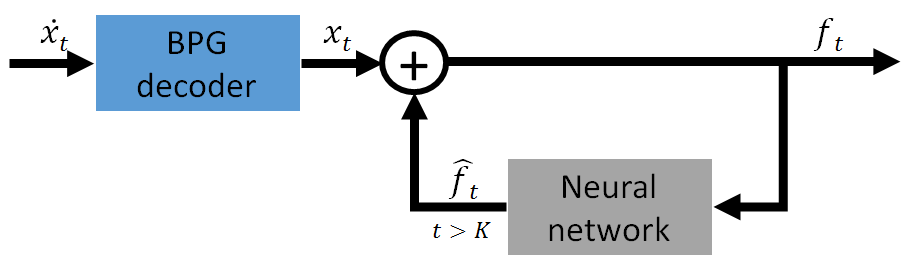}
	\caption{Block diagram of the decoder.}
	\label{fig:decoder}
\end{figure}

The decoder also runs a neural network with the same model parameters to predict the next frame given the previous $K$ decoded frames. The block diagram of the proposed video decoder is presented in Figure~\ref{fig:decoder}.
For the first $K$ frames, the decoder receives compressed full frames and the output of BPG decoder becomes the decoded image. After the first $K$ frames, the decoder receives the encoded frame differences, decodes them using the BPG decoder, and adds them to the predictions of the next frames generated by the neural network. In our experiments we used $K = 8$. Since the neural network model is sent to the decoder only once, only the encoded frame differences are transmitted and the proposed video codec is free from motion vector overhead.  

\chapter{Experimental Results}
\label{results}
This chapter presents experimental results to evaluate the performance of the proposed learned frame prediction method and its use for predictive video compression. We discuss the experimental setting and our evaluation methodology in Section~\ref{expset}. Learned frame prediction results are presented in Section~\ref{res:pred}. We provide the configuration parameters of the baseline video codecs used for comparison in Section~\ref{anchor}. Finally, Section~\ref{res:codec} presents our predictive video coding results and compare them with the results of the baseline video codecs.

\section{Experimental Setting and Evaluation Methodology}
\label{expset}
Our test dataset consists of 8 MPEG test sequences, shown in Table~\ref{table:test_videos}. We converted all videos into grayscale (Y-channel), which are stored in lossless png format.

\begin{table}[h]
\caption{MPEG test sequences.}
\centering
\begin{tabular}{|cccc|}
\hline
Sequence     & Pixels x Lines & Number of frames & Frames per sec. \\ \hline
Coastguard   & 352 x 288  & 300              & 25                \\
Container    & 352 x 288  & 300              & 25                \\
Football     & 352 x 240  & 125              & 25                \\
Foreman      & 352 x 288  & 300              & 25                \\
Garden       & 352 x 240  & 115              & 25                \\
Hall Monitor & 352 x 288  & 300              & 25                \\
Mobile       & 352 x 240  & 140              & 25                \\
Tennis       & 352 x 240  & 112              & 25                \\ \hline
\end{tabular}
\label{table:test_videos}
\end{table}

We first perform next frame prediction for all frames (except the first 8 frames) of these sequences using our two learned models, one trained with mean square loss only, and the other trained with a combination of mean square and adversarial losses. While the training is patch based, we perform full frame prediction during test time. The prediction results are evaluated in terms of the PSNR, defined by
\begin{equation}
PSNR = 10\log_{10}\left ( \frac{255^2}{L_{MS}} \right )
\label{psnr}
\end{equation}
where $L_{MS}$ denotes the mean square difference between learned frame prediction results and the original (ground-truth) frames as defined in Equation.~\ref{mse}. The PSNR of consecutive frame differences (without any prediction) and PSNR of predicted frames using $16 \times 16$ block motion compensation are used as baselines to compare the performance of our learned prediction models.

We then perform lossy predictive video compression on Y-channels of these test sequences using our two learned frame prediction models (instead of the standard motion compensation) in fixed quantization parameter (QP), i.e. variable bitrate (VBR) setting.  For comparison, we compress the same videos using x264 codec library \cite{x264} as a baseline. Since our networks perform sequential prediction and compression, we configured the x264 codec for both sequential (IPP...) and hierarchical coding as well as both CBR and VBR modes for fair comparison. In order to obtain PSNR vs. bitrate curves for all videos, we used integer QPs ranging from 25 to 35 in the VBR mode. For CBR coding using x264 codec, we employed the rate-control option. Employing BPG to encode the residual images, we also performed compression with our baseline frame prediction methods, 16x16 block motion compensation and frame difference.

In summary, we compare performance of the following predictive coding settings:
\vspace{-30pt}
\begin{itemize}
  \item Learned frame prediction trained with mean square loss only (LFP-MSE) \vspace{-10pt}
  \item Learned frame prediction trained with a combination of mean square and adversarial losses (LFP-GAN) \vspace{-10pt}
  \item Sequential (IPP...) x264 with variable bitrate (SEQ-VBR) \vspace{-10pt}
  \item Hierarchical x264 with variable bitrate (HIE-VBR) \vspace{-10pt}
  \item Sequential (IPP...) x264 with constant bitrate (SEQ-CBR) \vspace{-10pt}
  \item Hierarchical x264 with constant bitrate (HIE-CBR) \vspace{-8pt}
  \item 16x16 block motion compensation (MC) \vspace{-8pt}
  \item Frame difference (FD) \vspace{-8pt}
\end{itemize}

To evaluate the performance of these methods, we compute rate-distortion (PSNR vs. bitrate) curves for all methods and videos. We use the \textit{PSNR}, defined by Eqn.~\ref{psnr}, as the distortion metric, where $L_{MS}$ denotes the mean square error, as defined in Eqn.~\ref{mse}, between the decompressed video (after adding the encoded prediction error to the predicted frames) and the original (ground-truth) frames.

Since we employ the BPG codec to encode predictive frame differences in a fixed QP setting in our method, we compute the resulting average \textit{bitrate} per second for each video by multiplying the compressed file size for each difference image with the frame rate and then averaging the resulting bits/second values over the sequence. The output of x264 encoder is a single file. In order to compute its bitrate, we divide the file size by the number of frames and multiply the result with the frame rate.

\begin{figure}[h]
	\centering
	\includegraphics[scale=0.35]{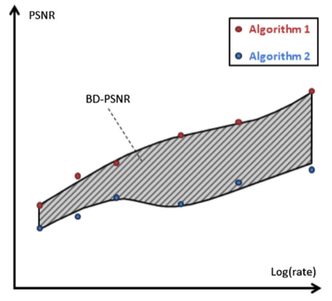}
	\caption{Computation of the Bjontegaard delta PSNR (BD-PSNR) metric \cite{bjontegaardfigure}.}
\label{fig:bjontegaard}
\end{figure}

We compare rate distortion curve of each method with that of the anchor method using the Bjontegaard delta (BD-PSNR) metric \cite{bjontegaard}, which measures the difference of areas between two PSNR vs. bitrate curves, as shown in Figure \ref{fig:bjontegaard}. The bitrates are expressed in logarithmic scale and continuous curves are formed by fitting a third order polynomial to available data points. The limits of integration are set as the higher one of the minimum bitrates and the lower one of the maximum bitrates. Area under the curve of the anchor method is subtracted from the area under the curve of the each method; hence, a positive value means the method performs better than the anchor.

\section{Frame Prediction Results}
\label{res:pred}
This section presents experimental results for both quantitative and qualitative evaluation of our two learned video frame prediction models: the LFP-MSE and LFP-GAN models without considering the effect of compression. That is, in this section all prediction results are obtained by using original (uncompressed) previous frames.

\begin{figure}[ht]
\begin{center}
\subfloat      {
\includegraphics[width=0.7\linewidth]{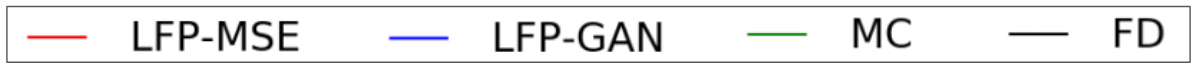}
}\\
\subfloat      {
\includegraphics[width=.45\linewidth]{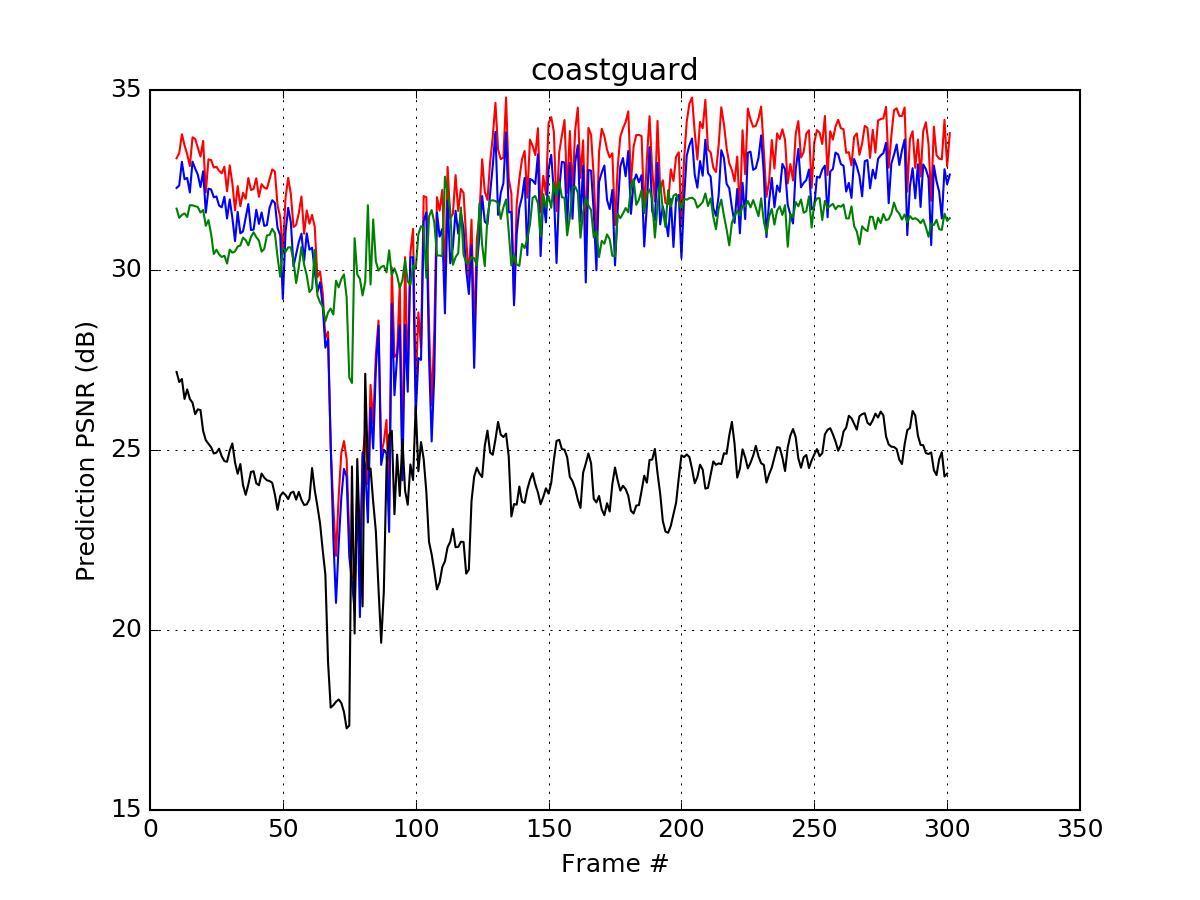}
}
\subfloat     {
\includegraphics[width=.45\linewidth]{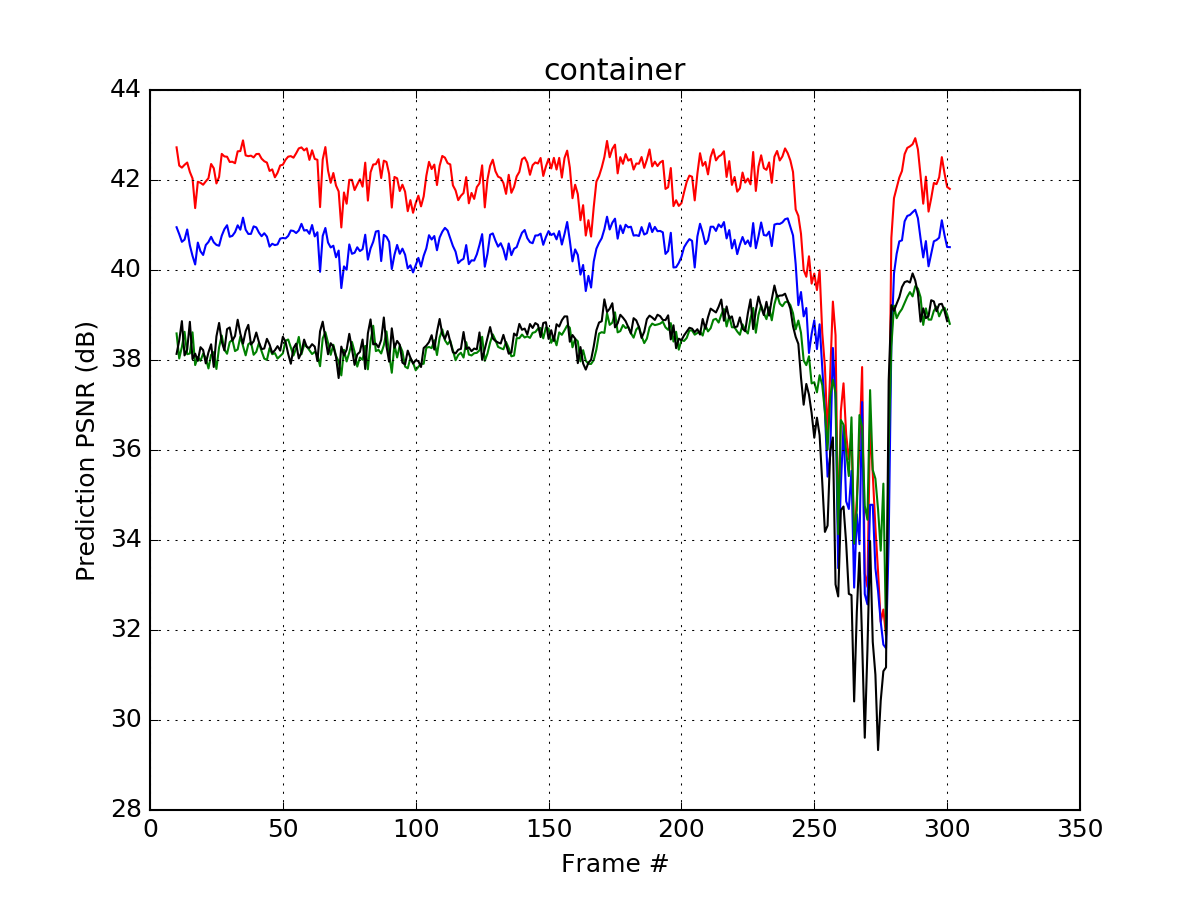}
}\\

\subfloat     {
\includegraphics[width=.45\linewidth]{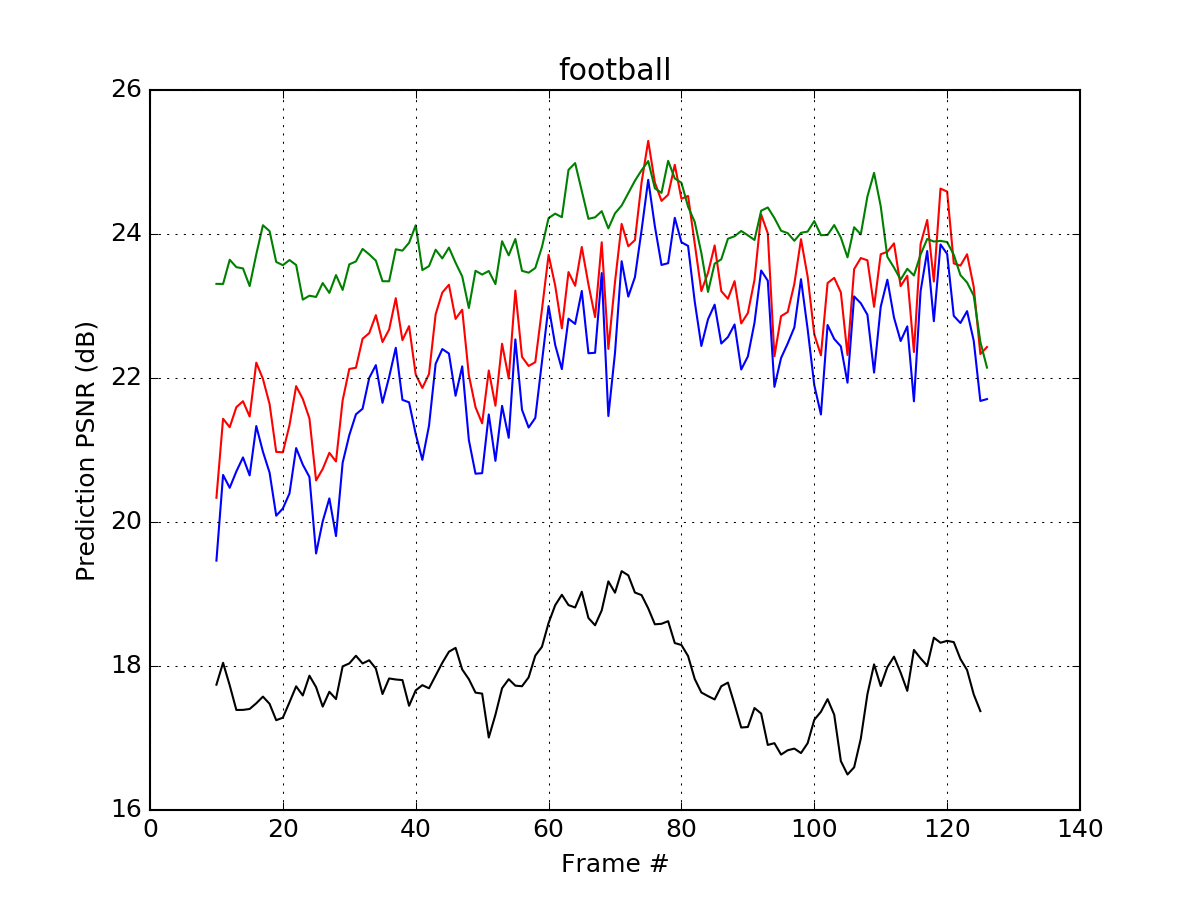}
}
\subfloat     {
\includegraphics[width=.45\linewidth]{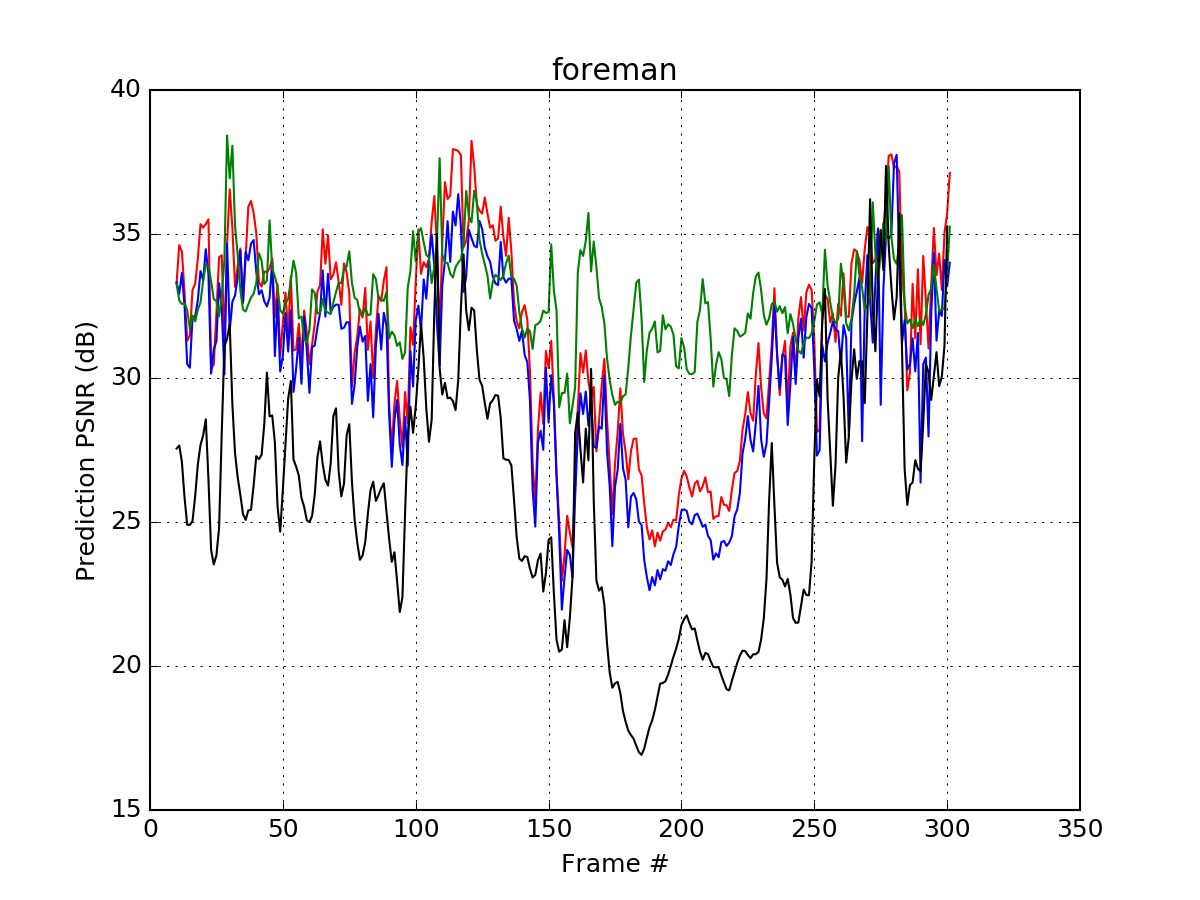}
}\quad

\subfloat     {
\includegraphics[width=.45\linewidth]{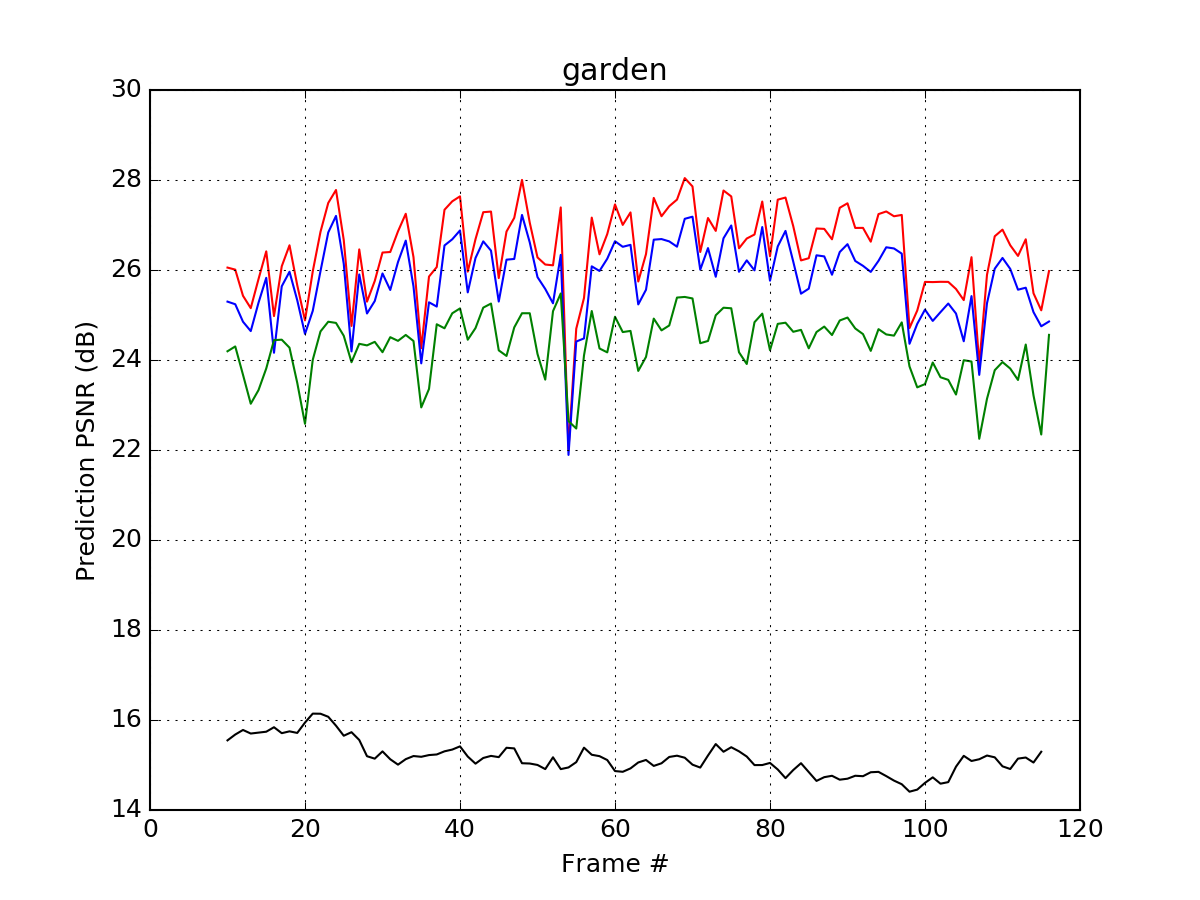}
}
\subfloat     {
\includegraphics[width=.45\linewidth]{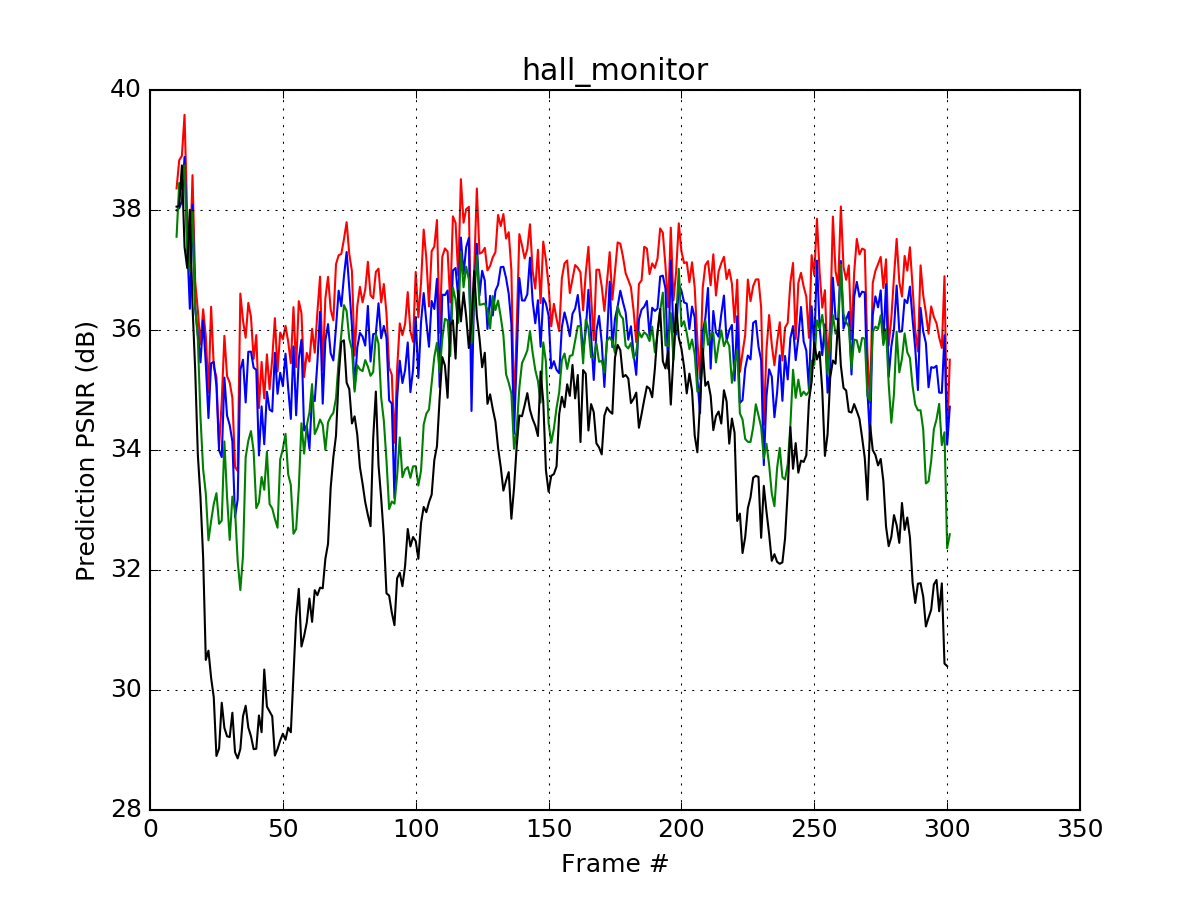}
}\\

\subfloat     {
\includegraphics[width=.45\linewidth]{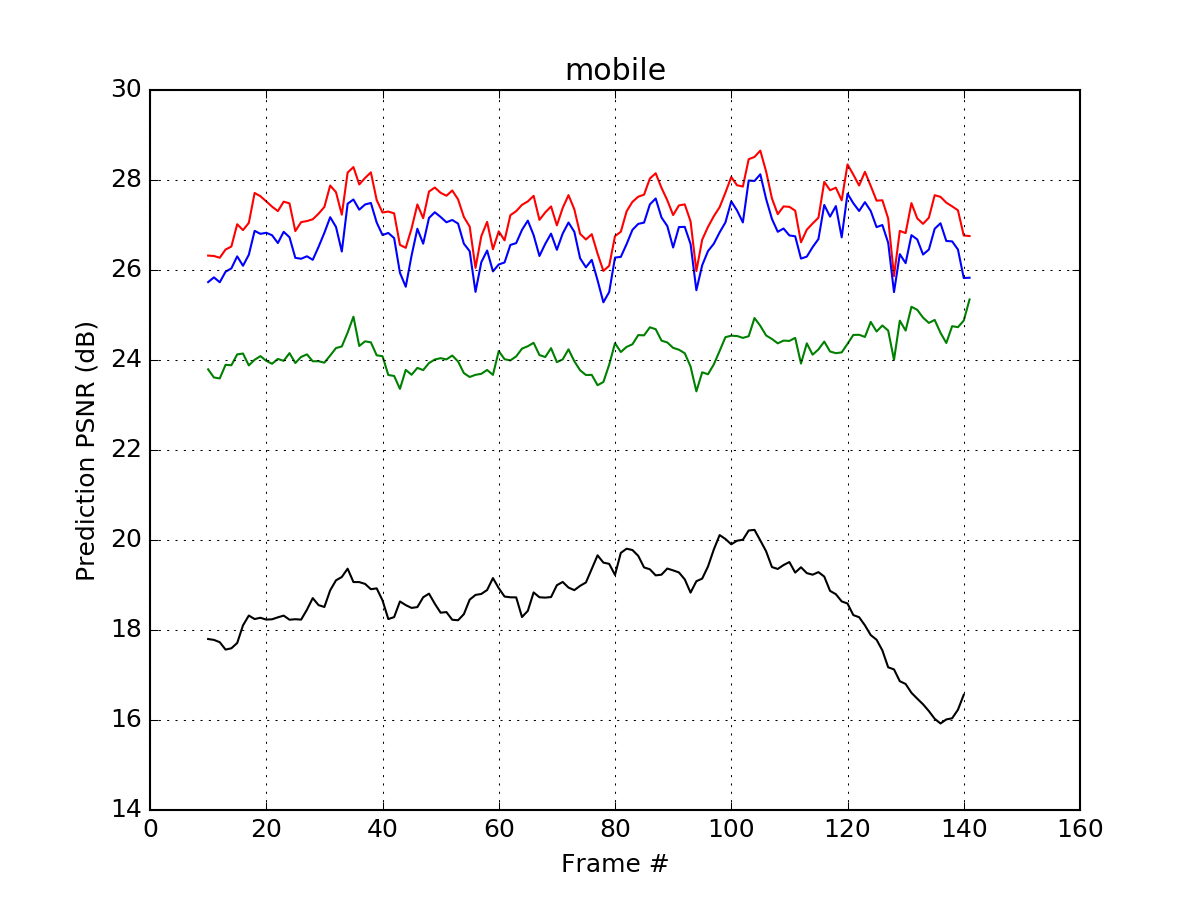}
}
\subfloat     {
\includegraphics[width=.45\linewidth]{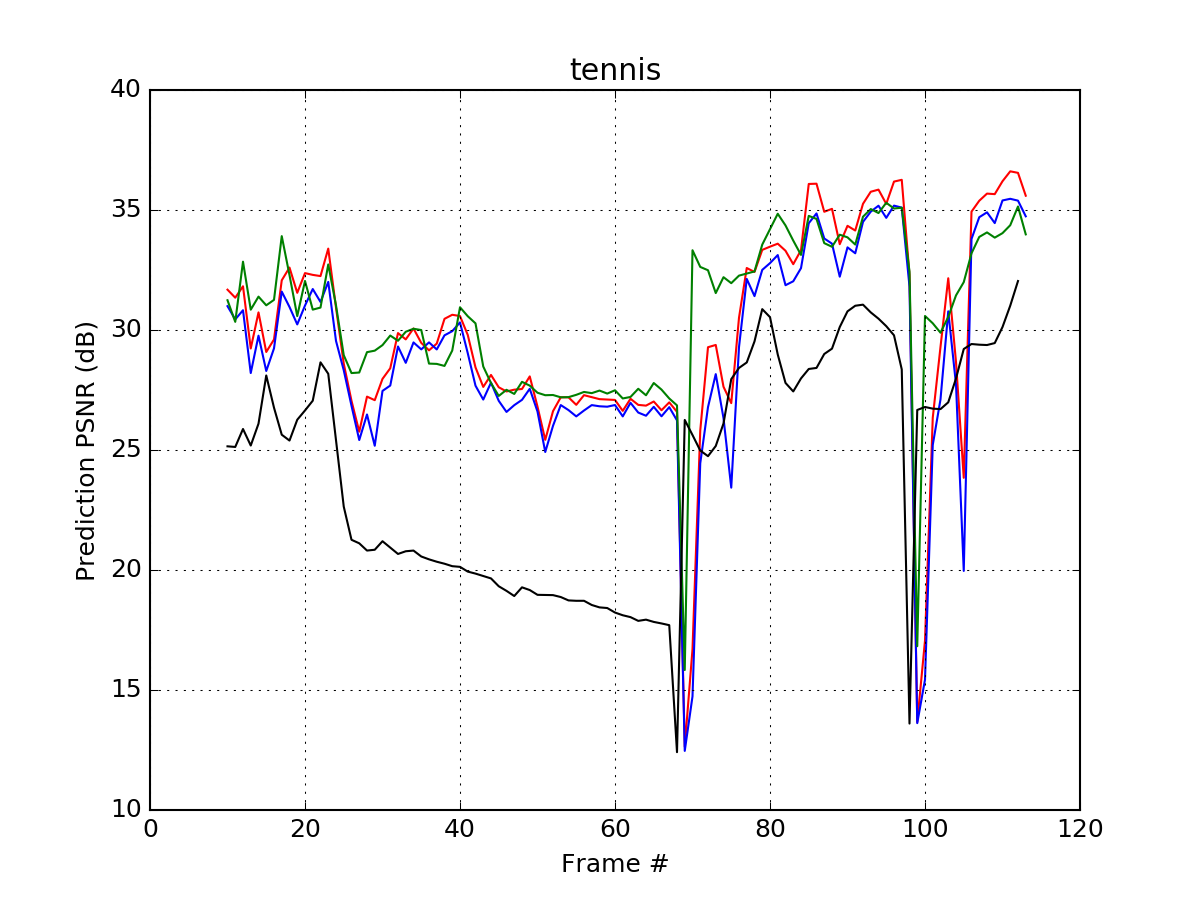}
}
\caption{PSNR of predicted frames vs. frame number for all test videos.}
\label{fig:prediction}
\end{center}
\end{figure}

We first perform quantitative evaluation in terms of prediction PSNR results by comparing each output frame obtained by our models (generator network) to the corresponding frame of the original video sequence (ground-truth). In Figure \ref{fig:prediction}, these values are plotted against the frame number for each video. We compare our methods to two baselines. The first baseline method is motion compensation (MC) with 16x16 blocks (patches), half pixel precision and exhaustive search with a motion vector limit of 31. The second baseline method is frame difference (FD), that is using the current frame as a prediction for the next frame.

As seen in Figure \ref{fig:prediction}, in each plot, both curves follow the same trends, with LFP-GAN producing inferior results. Since PSNR is inversely proportional to mean square error it is expected that training to minimize mean square error \textit{only} would produce the highest PSNR values. To train the LFP-GAN, we include the additional term of adversarial loss, hence we sacrifice mean square error for adversarial loss. As a result of this, the residual images for LFP-GAN have more information, namely higher entropy, so the compressed residual images have larger file sizes.

It can also be seen that, in container, a video with smooth surfaces and very slow motion, the gap between LFP-MSE and LFP-GAN is higher. This can be due to the fact that LFP-GAN aims at creating sharp patterns even when the original frame has smooth surfaces. Moreover, since the video has slow motion, LFP-MSE does not yield a high mean square error, hence LFP-GAN's attempt at reducing adversarial loss at the cost of increasing mean square loss becomes a dominating factor. 

We further observe that FD performs the worst for all videos. MC performs the best only at the presence of fast motion, such as the entire football video and parts of coastguard, foreman and tennis videos. Especially in football, a video with fast and complex motion, MC is capable of dealing with this complexity by relocating blocks of pixels independently. It is noted that MC algorithm has access to the target frame and it always relocates the block which yields the highest PSNR value. 

\begin{figure}[h]

\begin{center}
\vspace{0.8in}
\subfloat     {
\includegraphics[width=.48\linewidth]{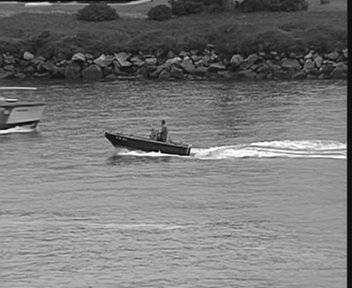}
}
\subfloat     {
\includegraphics[width=.48\linewidth]{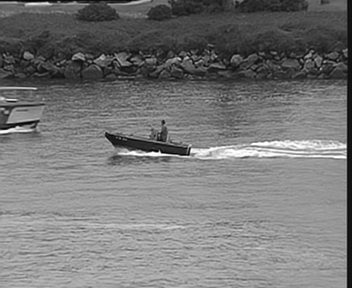}
} \\
(a) \hspace{2.7in} (b) \\
\subfloat     {
\includegraphics[width=.48\linewidth]{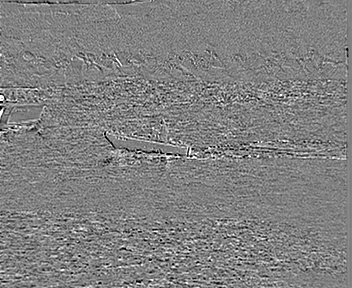}
}
\subfloat     {
\includegraphics[width=.48\linewidth]{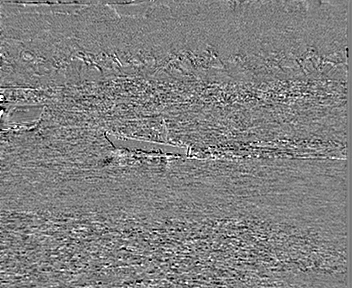}
} \\
(c) \hspace{2.7in} (d) \\
\caption{Coastguard. (a) Predicted frame using LFP-MSE method, (b) Predicted frame using LFP-GAN method, (c) Frame difference of LFP-MSE, (d) Frame difference of LFP-GAN. Difference values are scaled by 5 and level shifted by 128.}
\label{fig:coastguard}
\end{center}
\end{figure}

\begin{figure}[h]

\begin{center}
\vspace{0.8in}
\subfloat     {
\includegraphics[width=.48\linewidth]{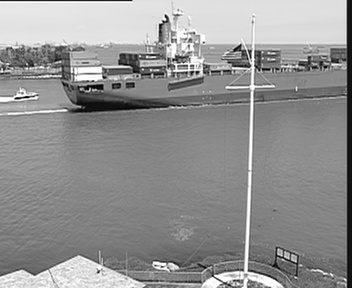}
}
\subfloat     {
\includegraphics[width=.48\linewidth]{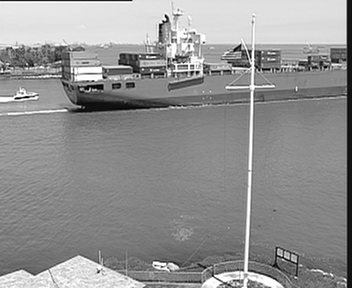}
} \\
(a) \hspace{2.7in} (b) \\
\subfloat     {
\includegraphics[width=.48\linewidth]{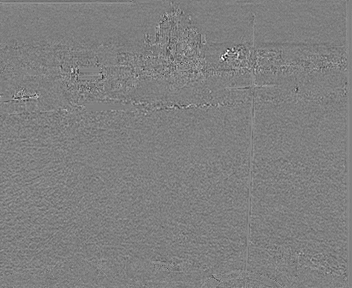}
}
\subfloat     {
\includegraphics[width=.48\linewidth]{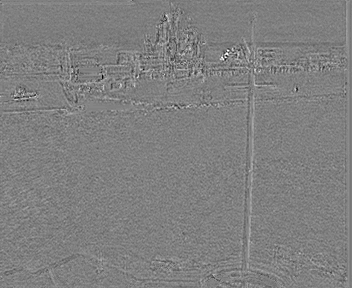}
} \\
(c) \hspace{2.7in} (d) \\
\caption{Container. (a) Predicted frame using LFP-MSE method, (b) Predicted frame using LFP-GAN method, (c) Frame difference of LFP-MSE, (d) Frame difference of LFP-GAN. Difference values are scaled by 5 and level shifted by 128.}
\label{fig:container}
\end{center}
\end{figure}

\begin{figure}[h]

\begin{center}
\vspace{0.8in}
\subfloat     {
\includegraphics[width=.48\linewidth]{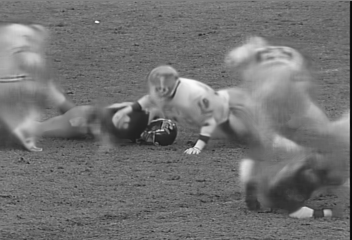}
}
\subfloat     {
\includegraphics[width=.48\linewidth]{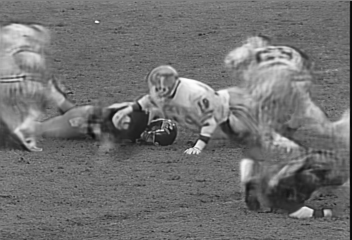}
} \\
(a) \hspace{2.7in} (b) \\
\subfloat     {
\includegraphics[width=.48\linewidth]{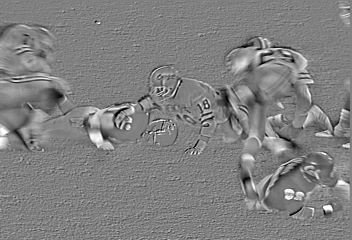}
}
\subfloat     {
\includegraphics[width=.48\linewidth]{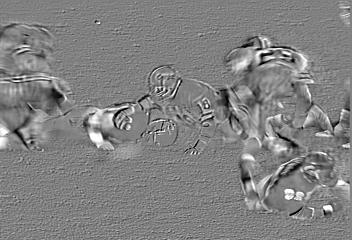}
} \\
(c) \hspace{2.7in} (d) \\
\caption{Football. (a) Predicted frame using LFP-MSE method, (b) Predicted frame using LFP-GAN method, (c) Frame difference of LFP-MSE, (d) Frame difference of LFP-GAN. Difference values are not scaled. They are level shifted by 128.}
\label{fig:football}
\end{center}
\end{figure}

\begin{figure}[h]

\begin{center}
\vspace{0.8in}
\subfloat     {
\includegraphics[width=.48\linewidth]{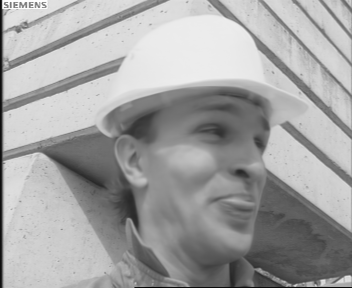}
}
\subfloat     {
\includegraphics[width=.48\linewidth]{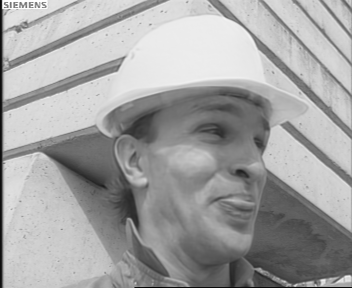}
} \\
(a) \hspace{2.7in} (b) \\
\subfloat     {
\includegraphics[width=.48\linewidth]{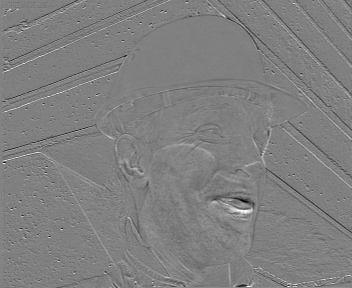}
}
\subfloat     {
\includegraphics[width=.48\linewidth]{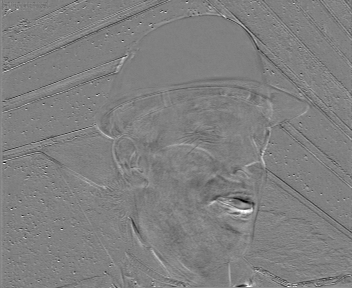}
} \\
(c) \hspace{2.7in} (d) \\
\caption{Foreman. (a) Predicted frame using LFP-MSE method, (b) Predicted frame using LFP-GAN method, (c) Frame difference of LFP-MSE, (d) Frame difference of LFP-GAN. Difference values are scaled by 2 and level shifted by 128.}
\label{fig:foreman}
\end{center}
\end{figure}

\begin{figure}[h]

\begin{center}
\vspace{0.8in}
\subfloat     {
\includegraphics[width=.48\linewidth]{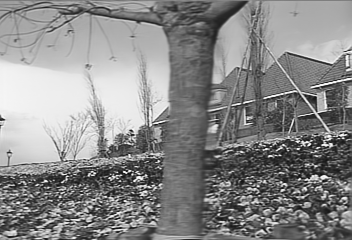}
}
\subfloat     {
\includegraphics[width=.48\linewidth]{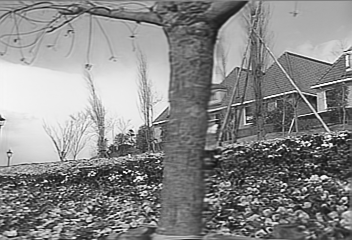}
} \\
(a) \hspace{2.7in} (b) \\
\subfloat     {
\includegraphics[width=.48\linewidth]{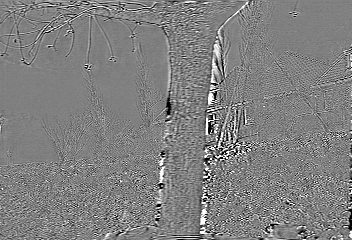}
}
\subfloat     {
\includegraphics[width=.48\linewidth]{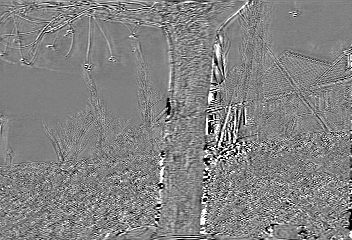}
} \\
(c) \hspace{2.7in} (d) \\
\caption{Garden. (a) Predicted frame using LFP-MSE method, (b) Predicted frame using LFP-GAN method, (c) Frame difference of LFP-MSE, (d) Frame difference of LFP-GAN. Difference values are scaled by 3 and level shifted by 128.}
\label{fig:garden}
\end{center}
\end{figure}

\begin{figure}[h]

\begin{center}
\vspace{0.8in}
\subfloat     {
\includegraphics[width=.48\linewidth]{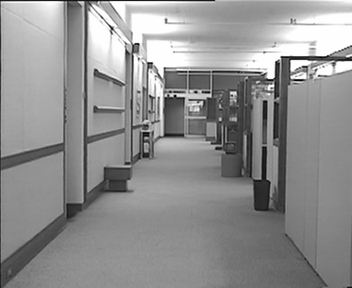}
}
\subfloat     {
\includegraphics[width=.48\linewidth]{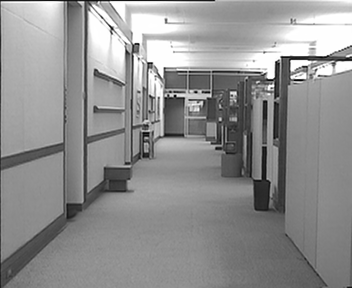}
} \\
(a) \hspace{2.7in} (b) \\
\subfloat     {
\includegraphics[width=.48\linewidth]{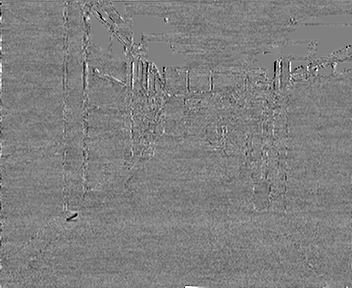}
}
\subfloat     {
\includegraphics[width=.48\linewidth]{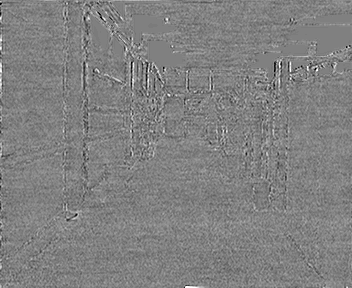}
} \\
(c) \hspace{2.7in} (d) \\
\caption{Hall Monitor. (a) Predicted frame using LFP-MSE method, (b) Predicted frame using LFP-GAN method, (c) Frame difference of LFP-MSE, (d) Frame difference of LFP-GAN. Difference values are scaled by 5 and level shifted by 128.}
\label{fig:hall_monitor}
\end{center}
\end{figure}

\begin{figure}[h]

\begin{center}
\vspace{0.8in}
\subfloat     {
\includegraphics[width=.48\linewidth]{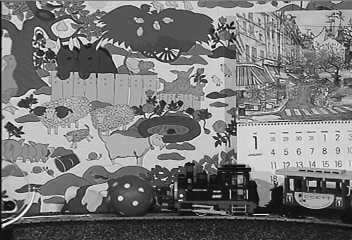}
}
\subfloat     {
\includegraphics[width=.48\linewidth]{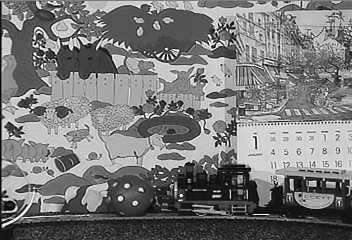}
} \\
(a) \hspace{2.7in} (b) \\
\subfloat     {
\includegraphics[width=.48\linewidth]{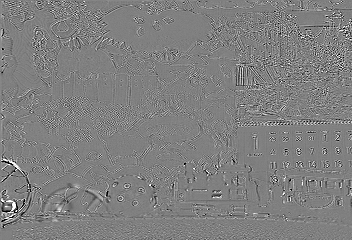}
}
\subfloat     {
\includegraphics[width=.48\linewidth]{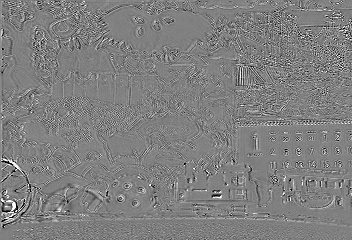}
} \\
(c) \hspace{2.7in} (d) \\
\caption{Mobile. (a) Predicted frame using LFP-MSE method, (b) Predicted frame using LFP-GAN method, (c) Frame difference of LFP-MSE, (d) Frame difference of LFP-GAN. Difference values are scaled by 2 and level shifted by 128.}
\label{fig:mobile}
\end{center}
\end{figure}

\begin{figure}[h]

\begin{center}
\vspace{0.8in}
\subfloat     {
\includegraphics[width=.48\linewidth]{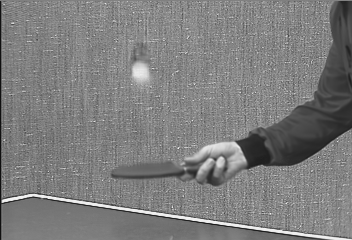}
}
\subfloat     {
\includegraphics[width=.48\linewidth]{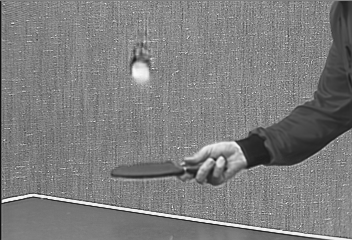}
} \\
(a) \hspace{2.7in} (b) \\
\subfloat     {
\includegraphics[width=.48\linewidth]{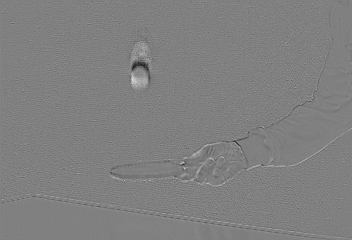}
}
\subfloat     {
\includegraphics[width=.48\linewidth]{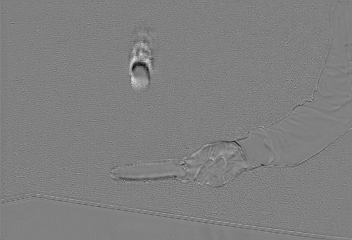}
} \\
(c) \hspace{2.7in} (d) \\
\caption{Tennis. (a) Predicted frame using LFP-MSE method, (b) Predicted frame using LFP-GAN method, (c) Frame difference of LFP-MSE, (d) Frame difference of LFP-GAN. Difference values are not scaled. They are level shifted by 128.}
\label{fig:tennis}
\end{center}
\end{figure}

Next, we present our qualitative results for the prediction of the 9th frames for each video. Through Figures \ref{fig:coastguard} to \ref{fig:tennis}, LFP-MSE (left) and LFP-GAN's (right) outputs can be seen on the upper row. Their difference with respect to the original frames are presented in the lower row. While the gray levels mean low difference, white and black levels denote high positive and negative difference. Despite its lower PSNR values, LFP-GAN produces sharper and more realistic looking images. In a detailed observation, it is seen that LFP-GAN produces higher error especially at the regions where LFP-MSE generates blurriness while LFP-GAN produces sharper surfaces. We can say that, in these regions, LFP-GAN takes a more stochastic approach.

\clearpage

\clearpage

\clearpage

\section{Baseline Video Codec Configurations}
\label{anchor}
We use the x264 codec supported by the FFmpeg library \cite{ffmpeg} as baseline for comparison with our compression results. FFmpeg provides the following options:

-i : Input file

-c:v : Video codec

-qp : Quantization parameter

-preset : Encoding preset (slower is better)

-r : Frame rate

-f : Output format

-g : Maximum keyframe interval

-keyint\_min : Minimum keyframe interval

-bf : Number of B-frames

-b:v : Video bitrate

-maxrate : Maximum bitrate

-minrate : Minimum bitrate

-bufsize : Rate control buffer size

-x264-params nal-hrd=cbr : Set network abstraction layer's (NAL) hypothetical reference decoder (HRD) to constant bitrate (CBR). Note that the output format needs to be ts, since mp4 format does not support this option \cite{ratecontrol}.

-x264-params force-cfr=1 : Force constant framerate timestamp generation

-pix\_fmt : Pixel format

\subsection{x264 Hierarchical Coding}
The standard (default) usage of the x264 codec, which usually yields the best results, is the hierarchical motion compensation (prediction) with variable bitrate (VBR) compression. The command line we used to encode the sequence "Coastguard" with QP value of 25 in HIE-VBR mode is provided as an example.

{\fontfamily{qcr}\selectfont
ffmpeg -i coastguard/\%3d.png -c:v libx264rgb -qp 25\\ -preset veryslow -r 25 -f mp4 coastguard.mp4
}

Alternatively, we can perform hierarchical prediction in constant bitrate (CBR) mode at a prespecified bitrate. The command line for x264 default CBR (HIE-CBR) coding with the bitrate of 1000k and buffer size of 2000k is provided as an example. 

{\fontfamily{qcr}\selectfont
ffmpeg -i coastguard/\%3d.png -c:v libx264rgb\\ -x264-params "nal-hrd=cbr:force-cfr=1" -b:v 1000k\\ -maxrate 1000k -minrate 1000k -bufsize 2000k\\ -preset veryslow -r 25 coastguard.ts
}

Note that FFmpeg documentation recommends a buffer size of 1-2 times the maximum bitrate \cite{streaming}.

\subsection{x264 Sequential Coding}
Because our learned models work sequentially, we also used x264 coding with sequential motion prediction, i.e., using a single I frame followed by P frames only. 

The command line for coding using x264 sequential VBR (SEQ-VBR) mode is

{\fontfamily{qcr}\selectfont
ffmpeg -i coastguard/\%3d.png -c:v libx264rgb -g 1000\\ -keyint\_min 1000 -bf 0 -qp 25 -preset veryslow -r 25\\ -f mp4 coastguard.mp4
}

Since the longest test video consists of 300 frames, by selecting 1000 as the minimum and maximum keyframe interval and zero as the B-frame number, we guarantee sequential motion estimation. This is verified through FFmpeg's output log.

Finally, the command line for x264 sequential CBR (SEQ-CBR) is as follows:

{\fontfamily{qcr}\selectfont
ffmpeg -i coastguard/\%3d.png -c:v libx264rgb -g 1000\\ -x264-params "nal-hrd=cbr:force-cfr=1" -b:v 1000k\\ -maxrate 1000k -minrate 1000k -bufsize 2000k -r 25\\ -preset veryslow -keyint\_min 1000 -bf 0 coastguard.ts
}

For all coding modes, after each command line, the following command line is used to decode the video into grayscale png frames:

{\fontfamily{qcr}\selectfont
ffmpeg -i coastguard.mp4 -pix\_fmt gray\\ coastguard\_decoded/\%3d.png
}

\section{Predictive Video Compression Results}
\label{res:codec}
This section evaluates the use of learned prediction models (instead of standard motion compensation) for predictive video compression. Naturally, in this setting all frame prediction results are obtained by using compressed/decompressed previous frames, since frame prediction must also be duplicated at the receiver. 

\begin{figure}[h]

\begin{center}
\subfloat     {
\includegraphics[width=.45\linewidth]{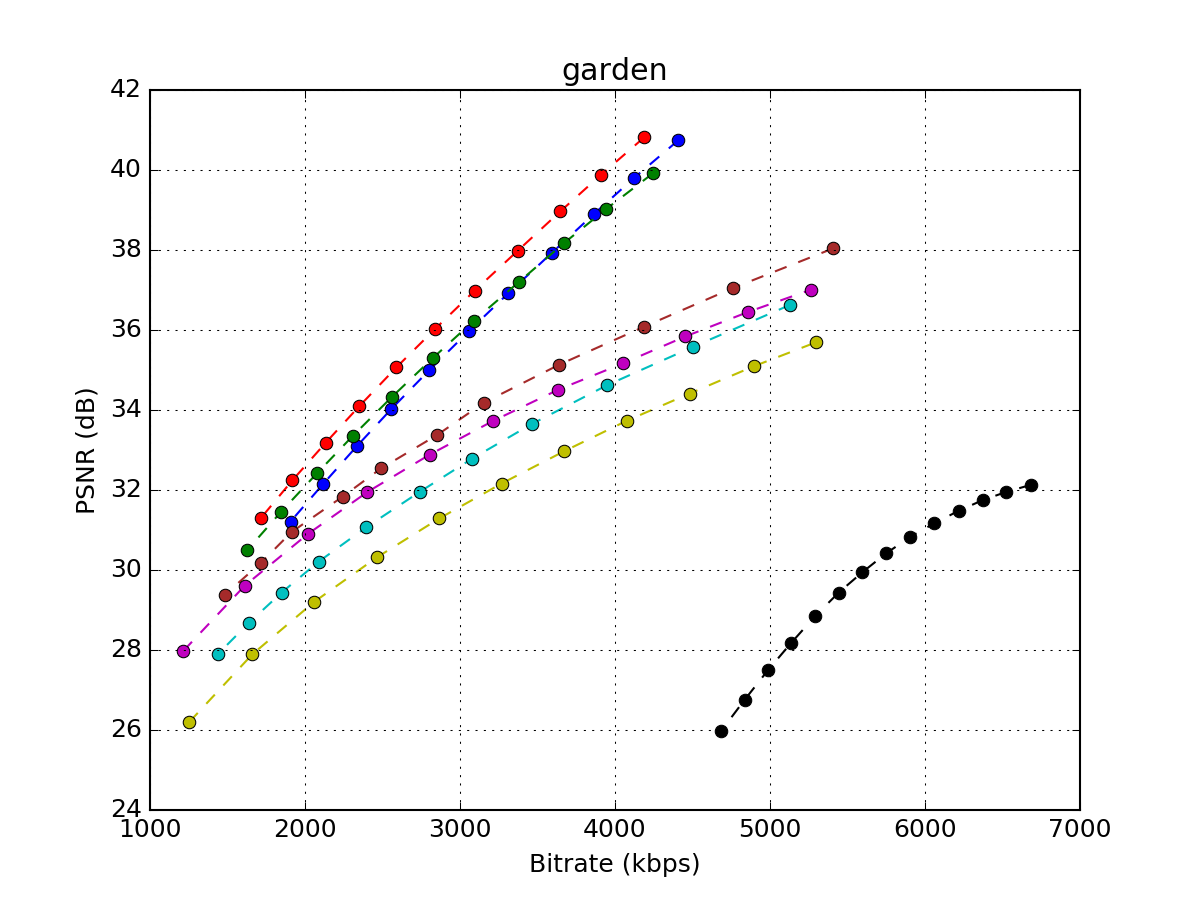}
}\quad
\subfloat     {
\includegraphics[width=.45\linewidth]{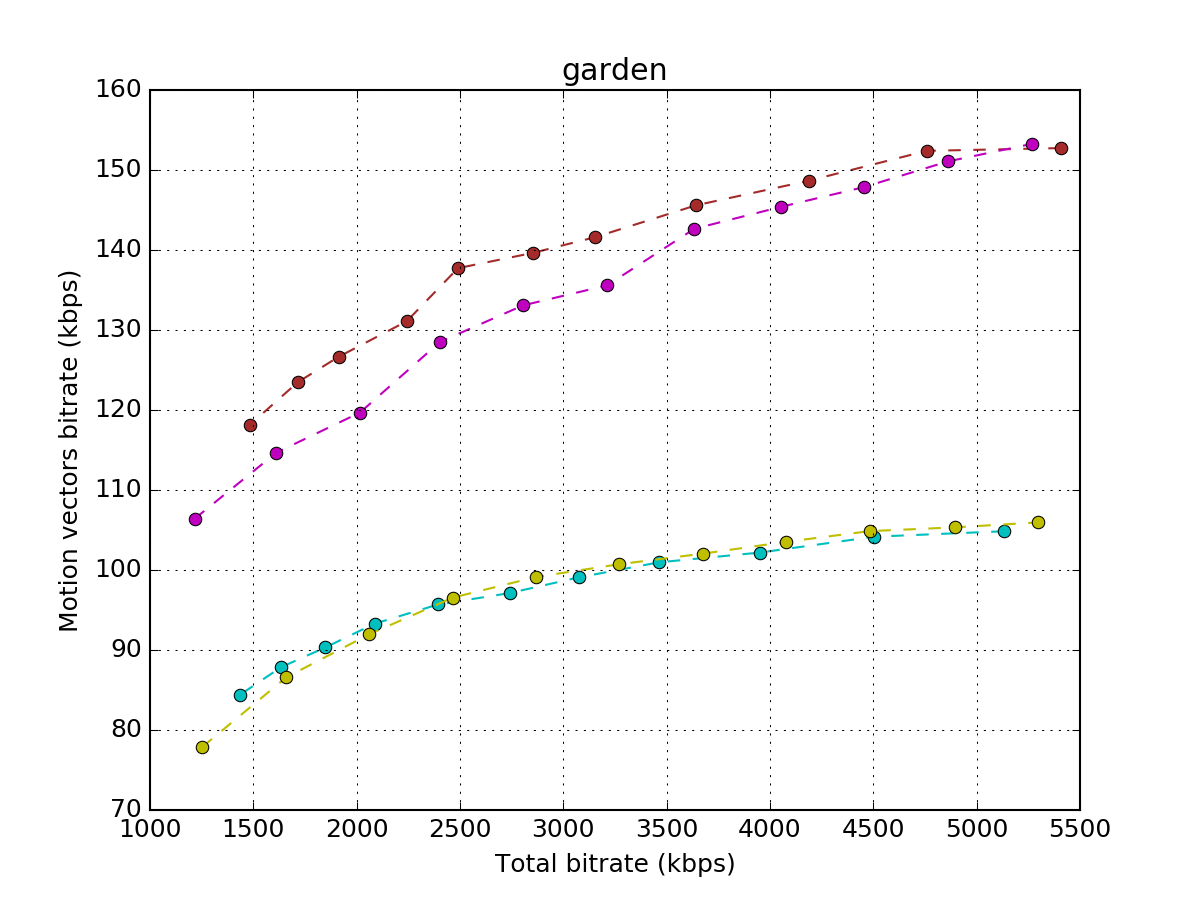}
}\\
\subfloat     {
\includegraphics[width=.45\linewidth]{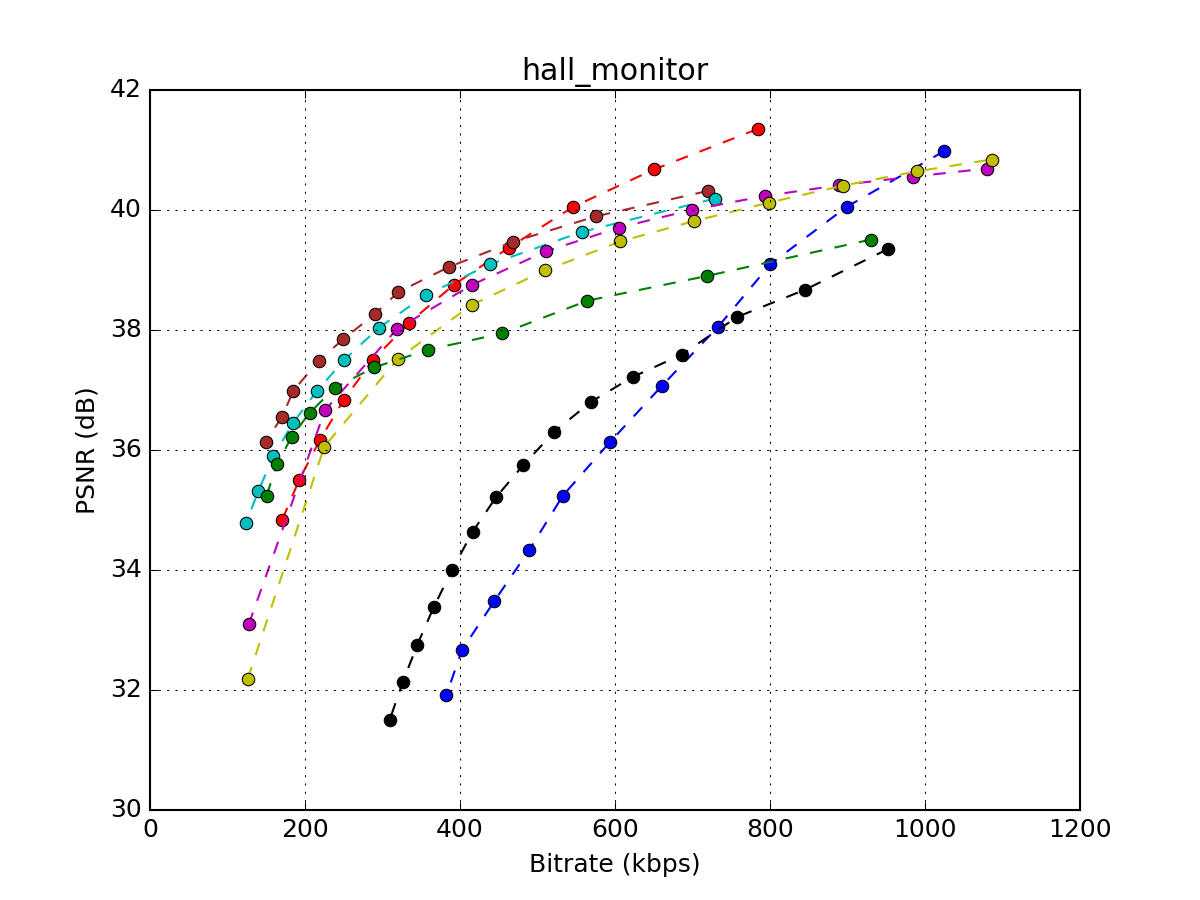}
}\quad
\subfloat     {
\includegraphics[width=.45\linewidth]{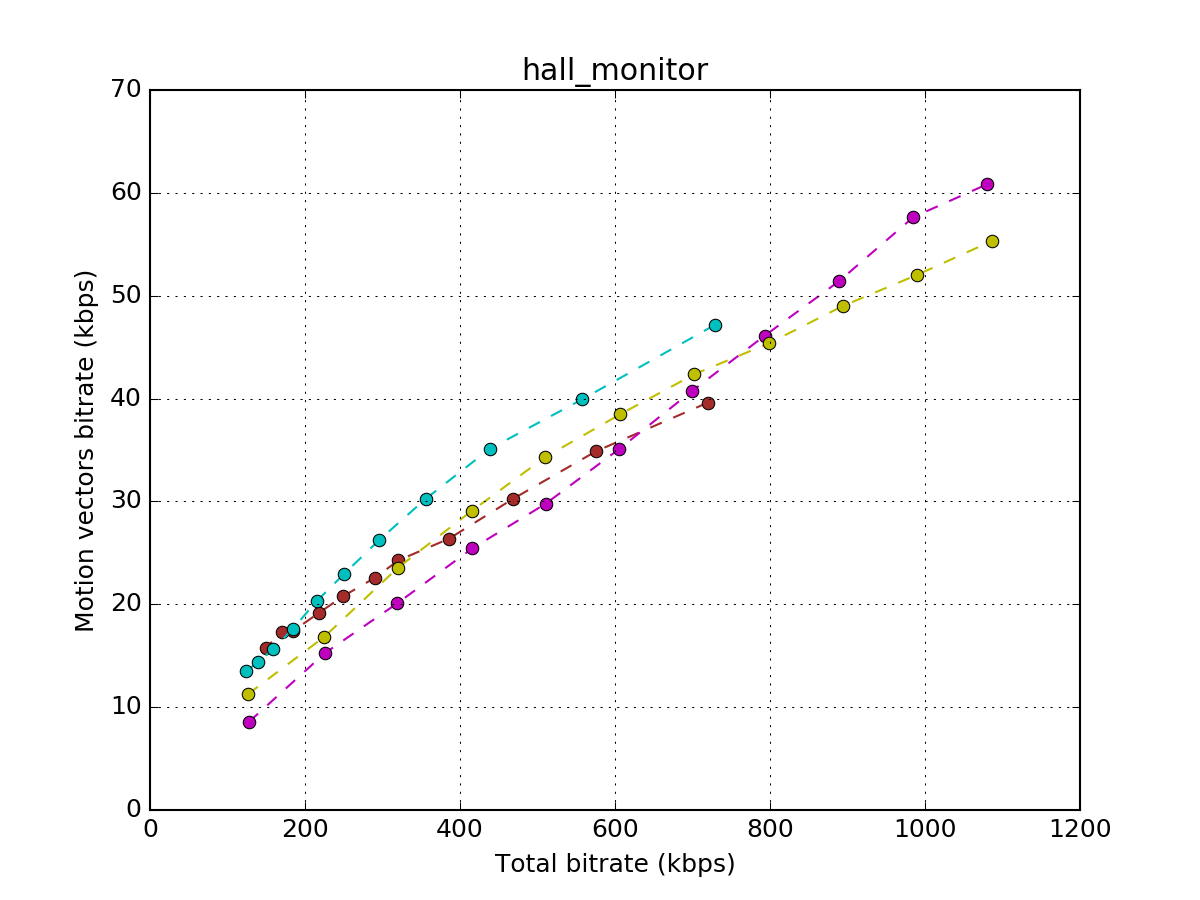}
}\\
\subfloat     {
\includegraphics[width=.45\linewidth]{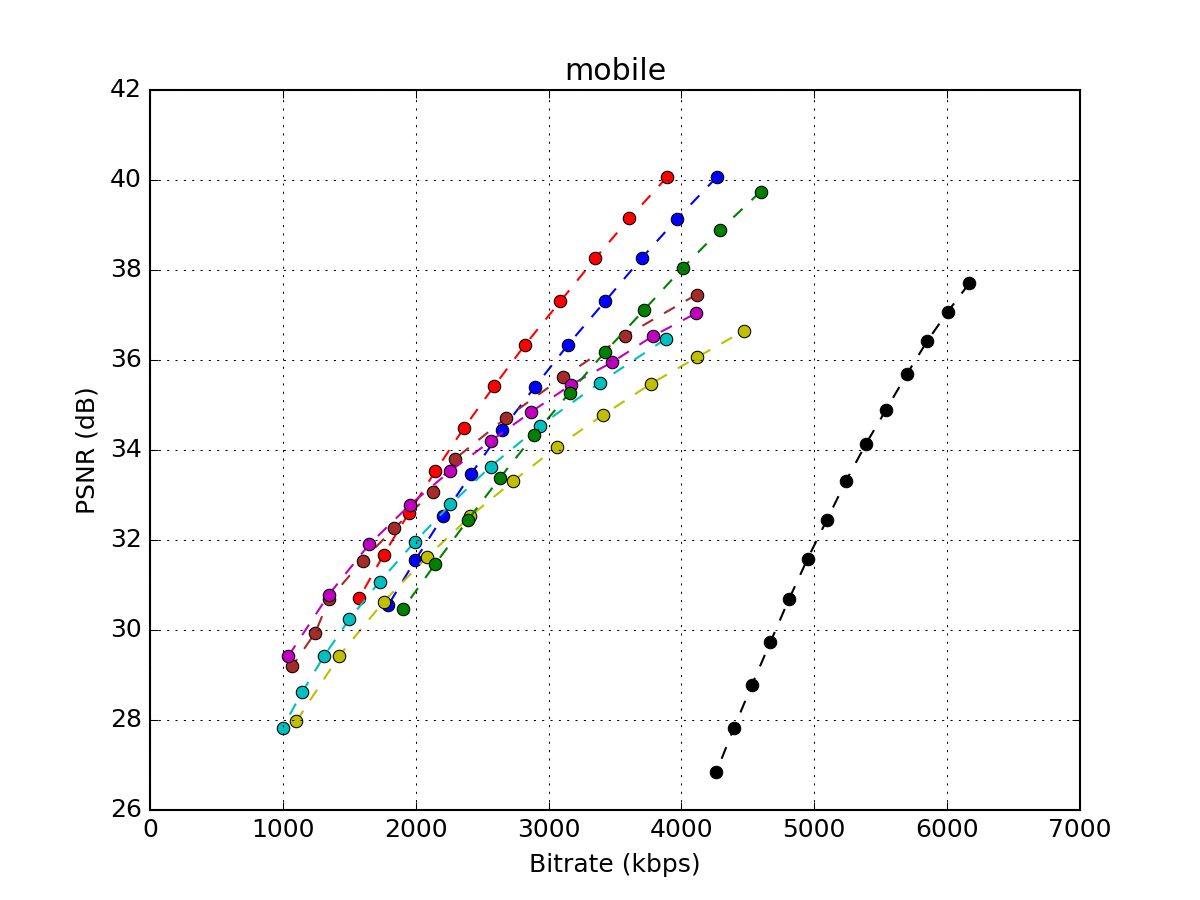}
}\quad
\subfloat     {
\includegraphics[width=.45\linewidth]{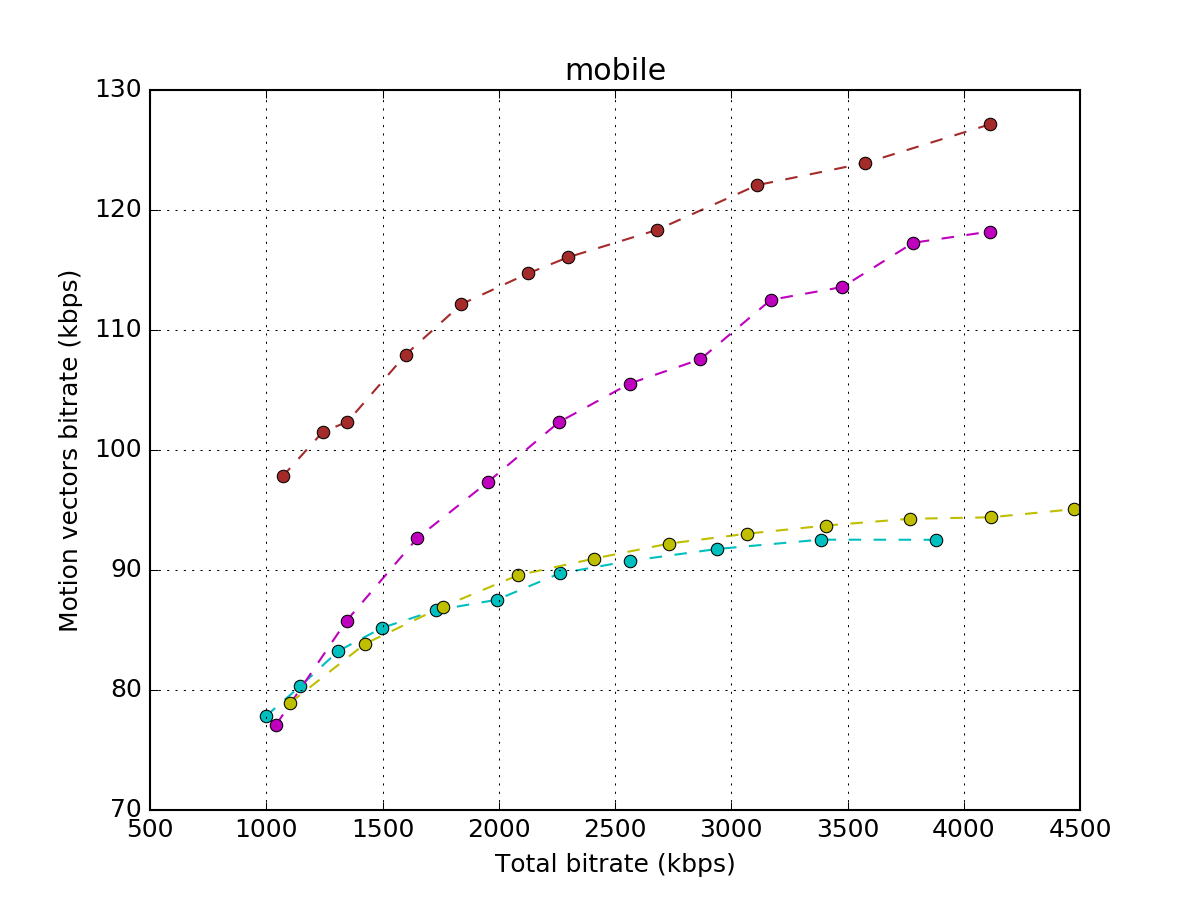}
}\\
\subfloat     {
\includegraphics[width=.45\linewidth]{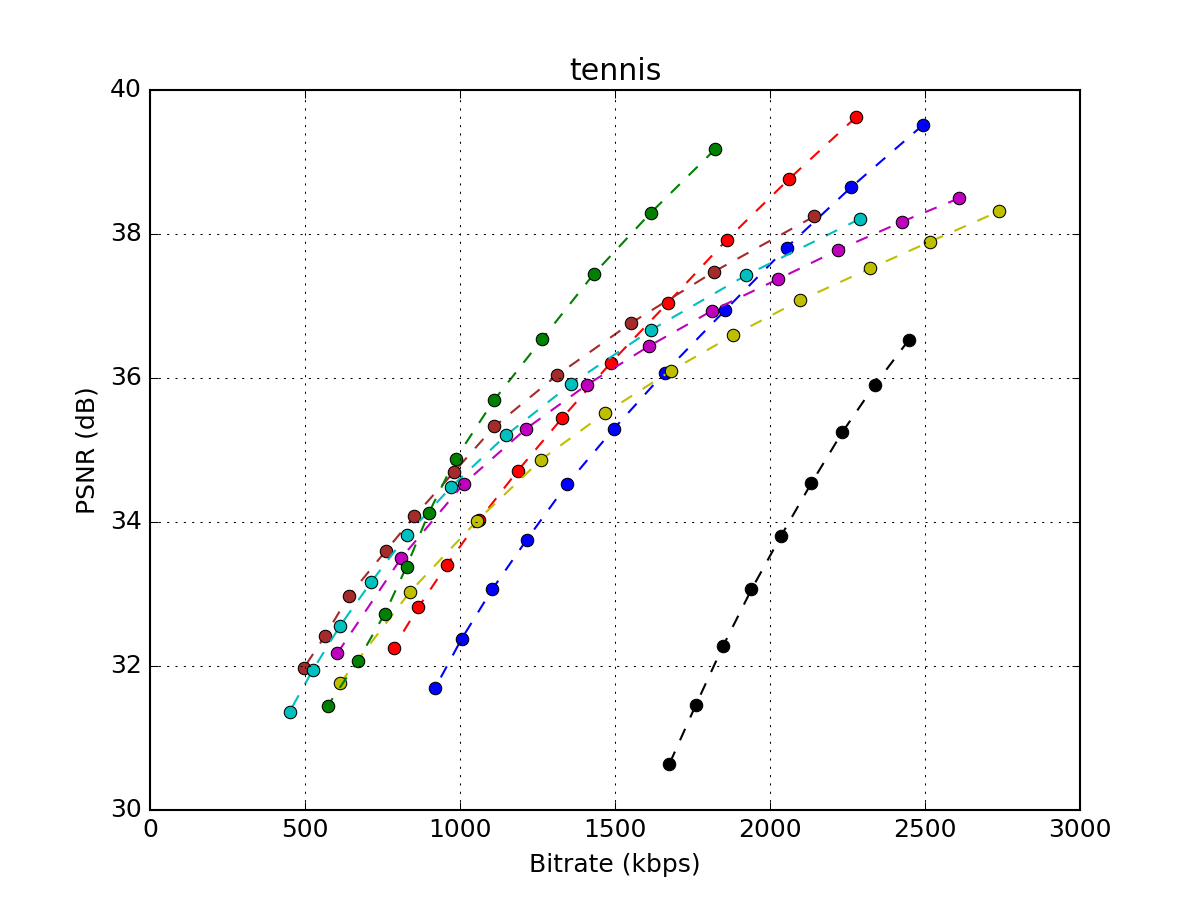}
}\quad
\subfloat     {
\includegraphics[width=.45\linewidth]{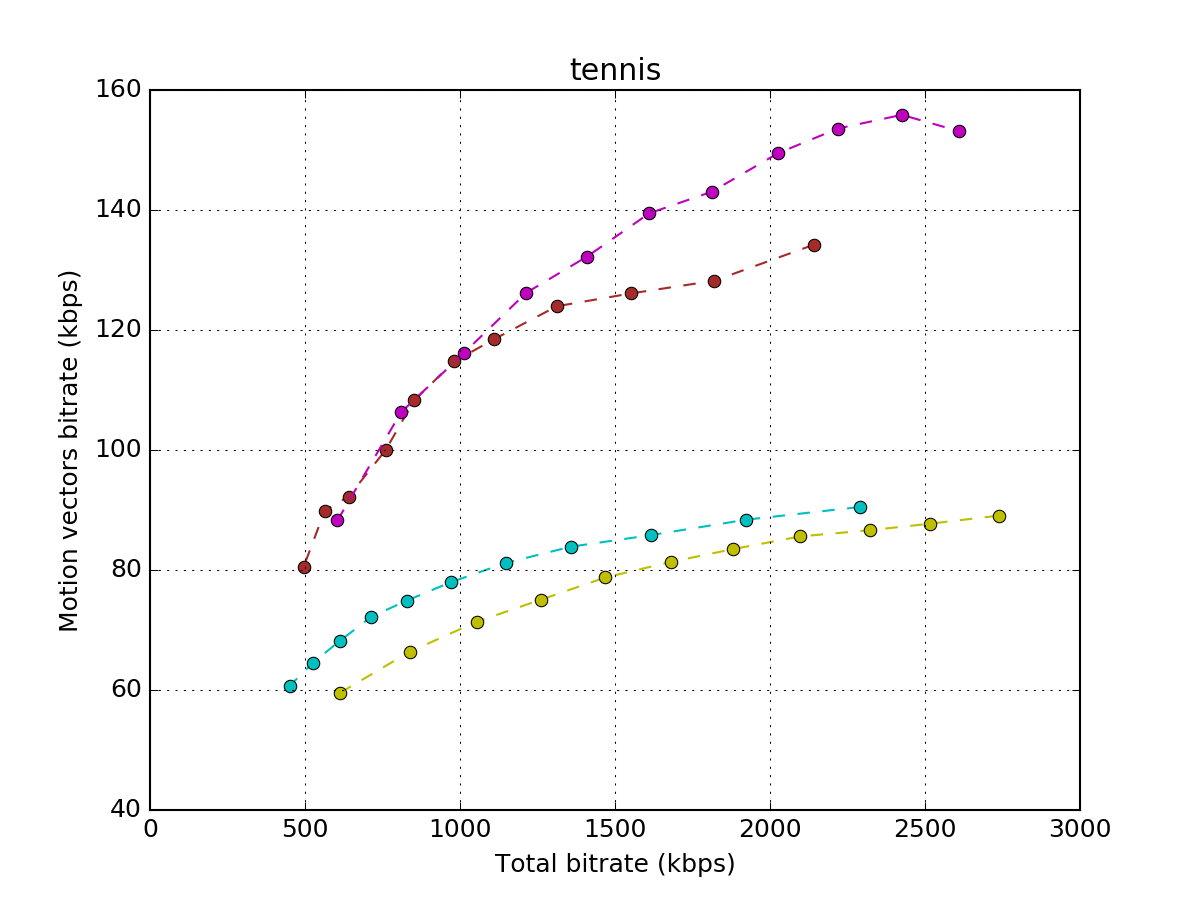}
}
\caption{Compression performance on test videos. Left: PSNR vs. bitrate. Right: Motion vector bitrate vs. total bitrate}
\label{fig:compression2}
\end{center}
\end{figure}

Using our neural networks, we present our compression performance compared to the x264 codec, MC and FD. In MC and FD, BPG is used to encode the residual image, just like in LFP methods. Only for x264, the motion vectors are extracted from the encoded videos, using the software \textit{mpegflow} \cite{mpegflow}. Bitrate of the motion vectors is obtained using their entropy. Note that there are no motion vectors for compression with neural networks. In Figure \ref{fig:compression2} rate-PSNR curves can be seen on the left. On the right we display the bitrate of the motion vectors of x264 codec versus total bitrate.

It is seen that in videos with high motion content, methods using BPG with LFP or 16x16 MC outperform x264. One of the reasons can be that, in these videos, residual images have more content, meaning higher entropy, and BPG is doing such a better job compressing these residual images that advanced motion compensation method of x264 is not sufficient to outperform the methods using BPG.

In videos with slow motion, the entropies of the residual images are low, so the superior motion compensation capability of x264 becomes an advantage over 16x16 MC. Also, for these videos, LFP-GAN performs much worse, as explained in Section~\ref{res:pred}. However, for all videos, our method of LFP-MSE remains competitive. 

In almost all videos, FD performs the worst. Here, the efficiency of BPG is not sufficient to compensate for the high entropy residual images.

Finally, to summarize our results, we show the Bjontegaard delta values calculated against the anchor method, in Table~\ref{table:bjontegaard}. Since our models work sequentially with no limitation on the bitrate, we have chosen sequential x264 codec with variable bitrate (SEQ-VBR), as the anchor method for a fair comparison.

It is observed that even though it generates sharper and more realistic images, GAN training worsens compression performance for all videos. Our method LFP-MSE, outperforms the anchor in 5 videos out of 8. In 2 of the 3 videos that LFP-MSE is outperformed by the anchor, the margin is very small. Furthermore in 4 videos, it outperforms HIE-VBR, the best possible configuration in x264 codec. 

\clearpage

\begin{figure}[h]

\begin{center}
\subfloat      {
\includegraphics[width=0.84\linewidth]{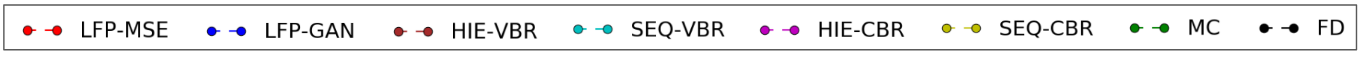}
}\\
\subfloat      {
\includegraphics[width=.45\linewidth]{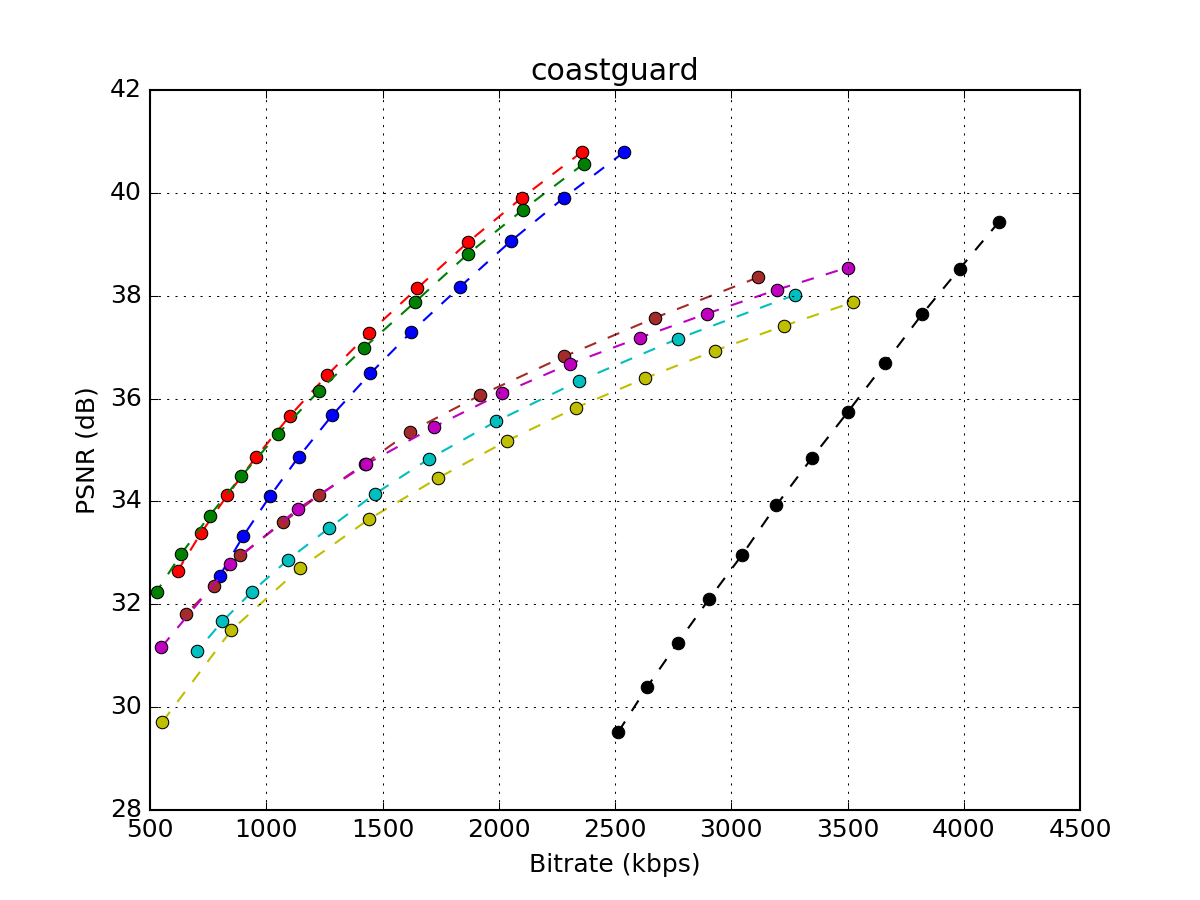}
}\quad
\subfloat     {
\includegraphics[width=.45\linewidth]{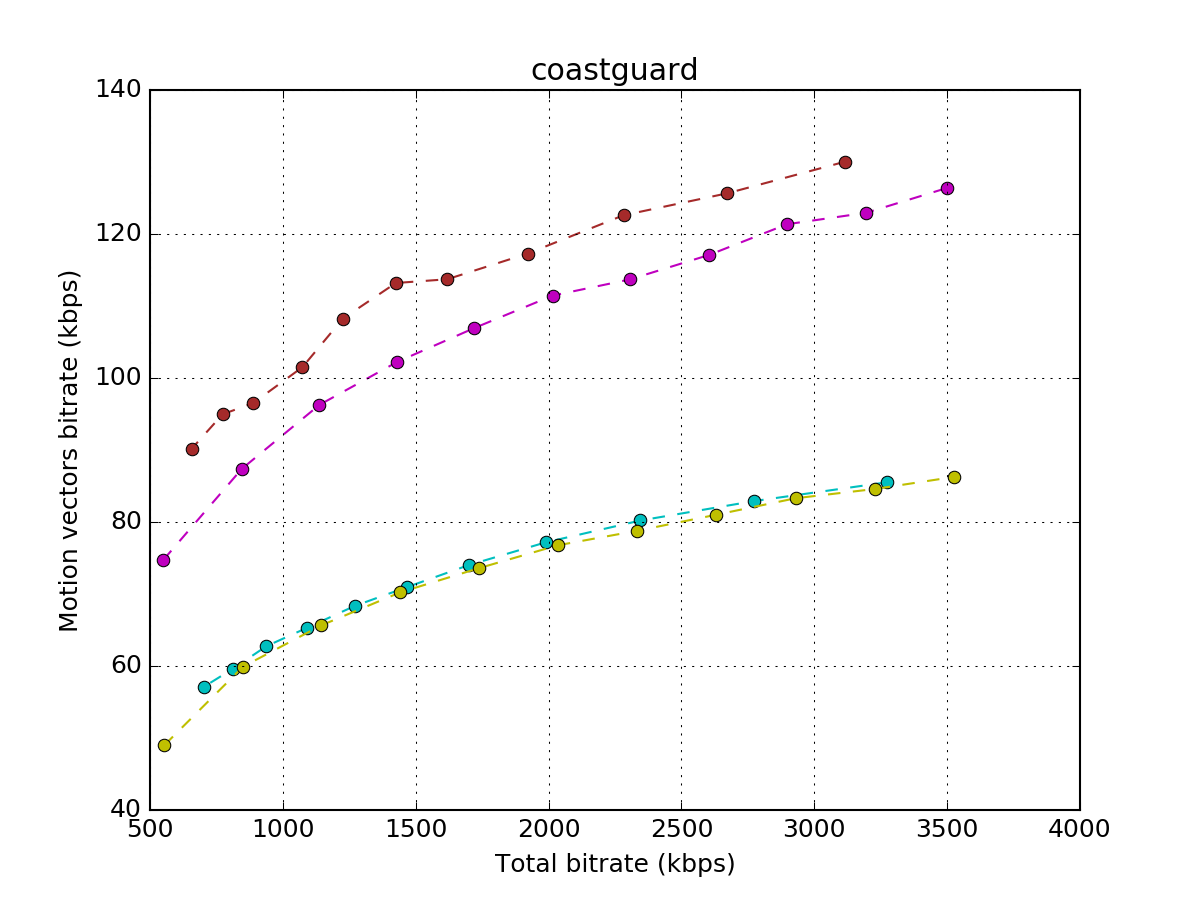}
}\\
\subfloat     {
\includegraphics[width=.45\linewidth]{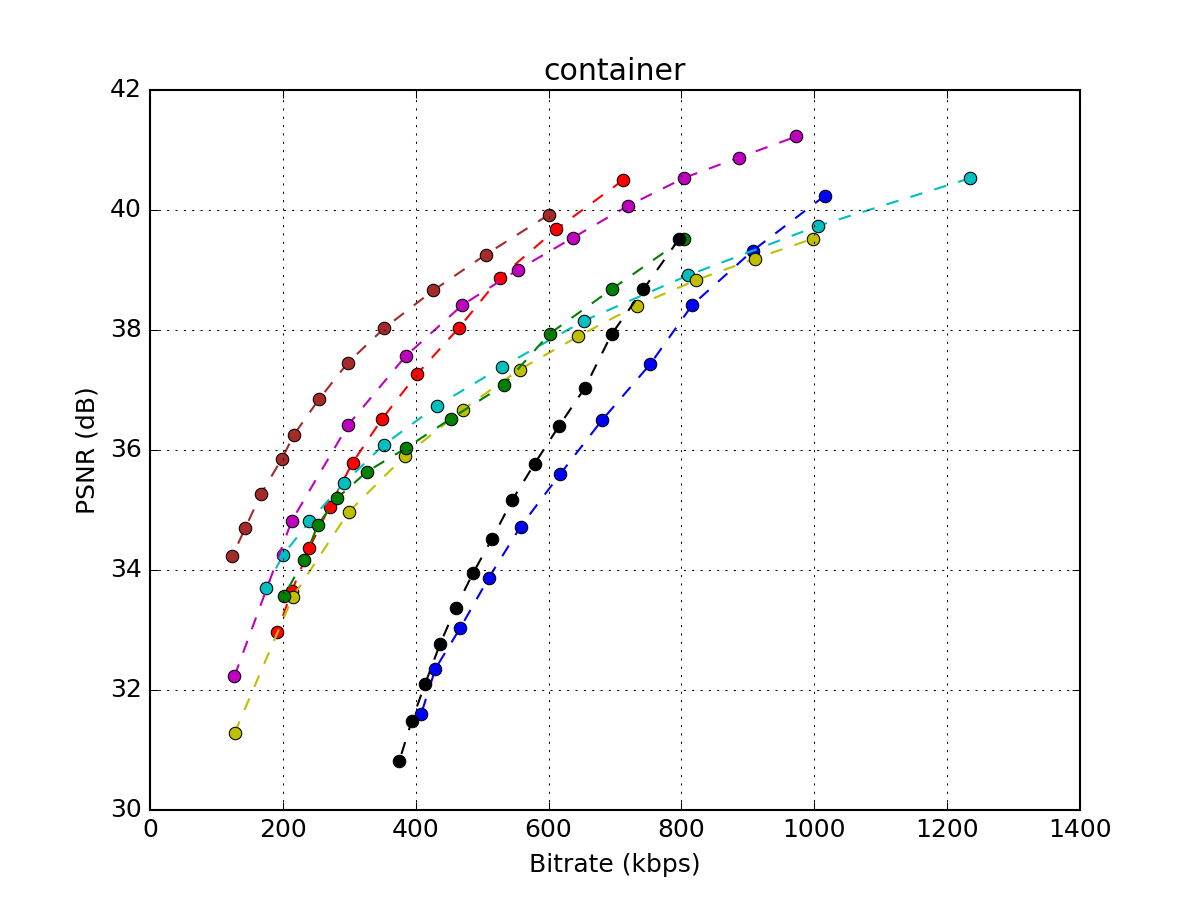}
}\quad
\subfloat     {
\includegraphics[width=.45\linewidth]{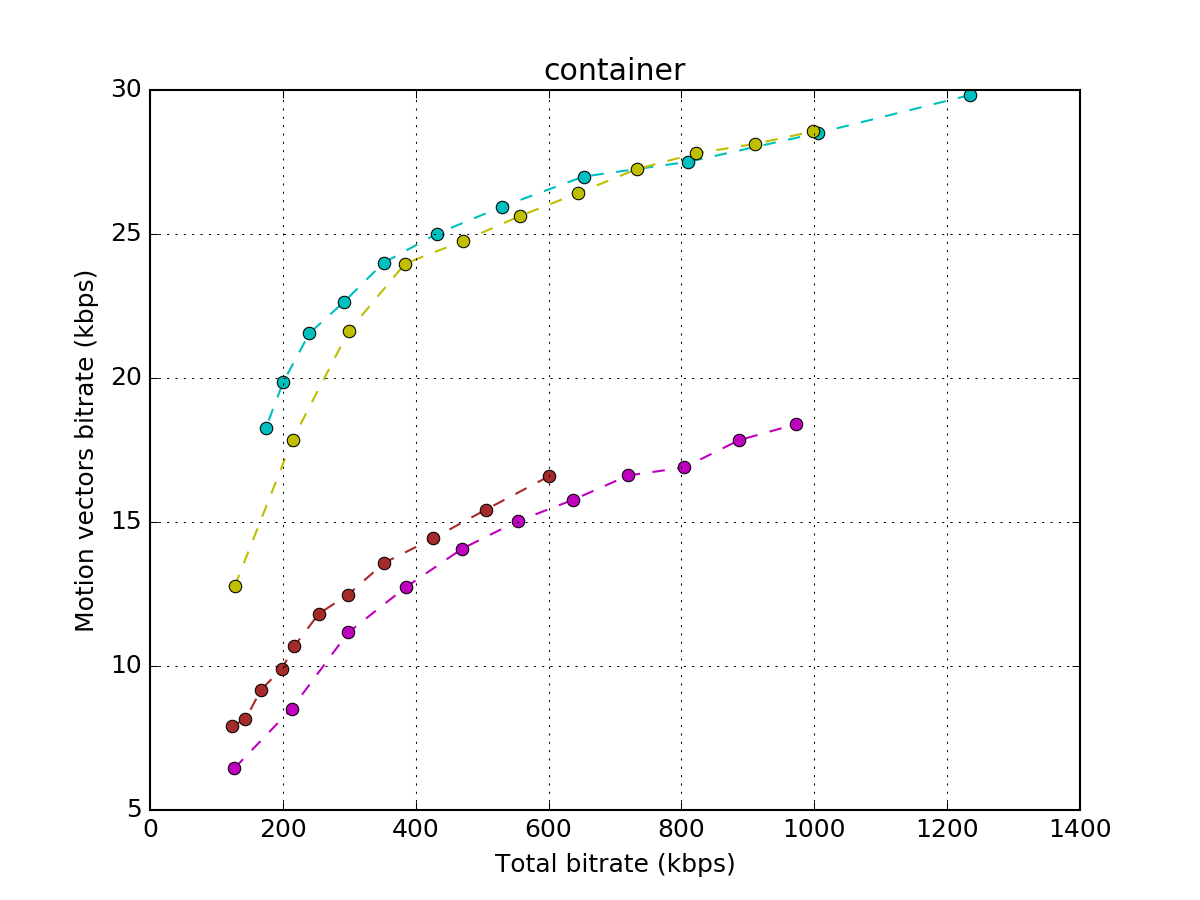}
}\\
\subfloat     {
\includegraphics[width=.45\linewidth]{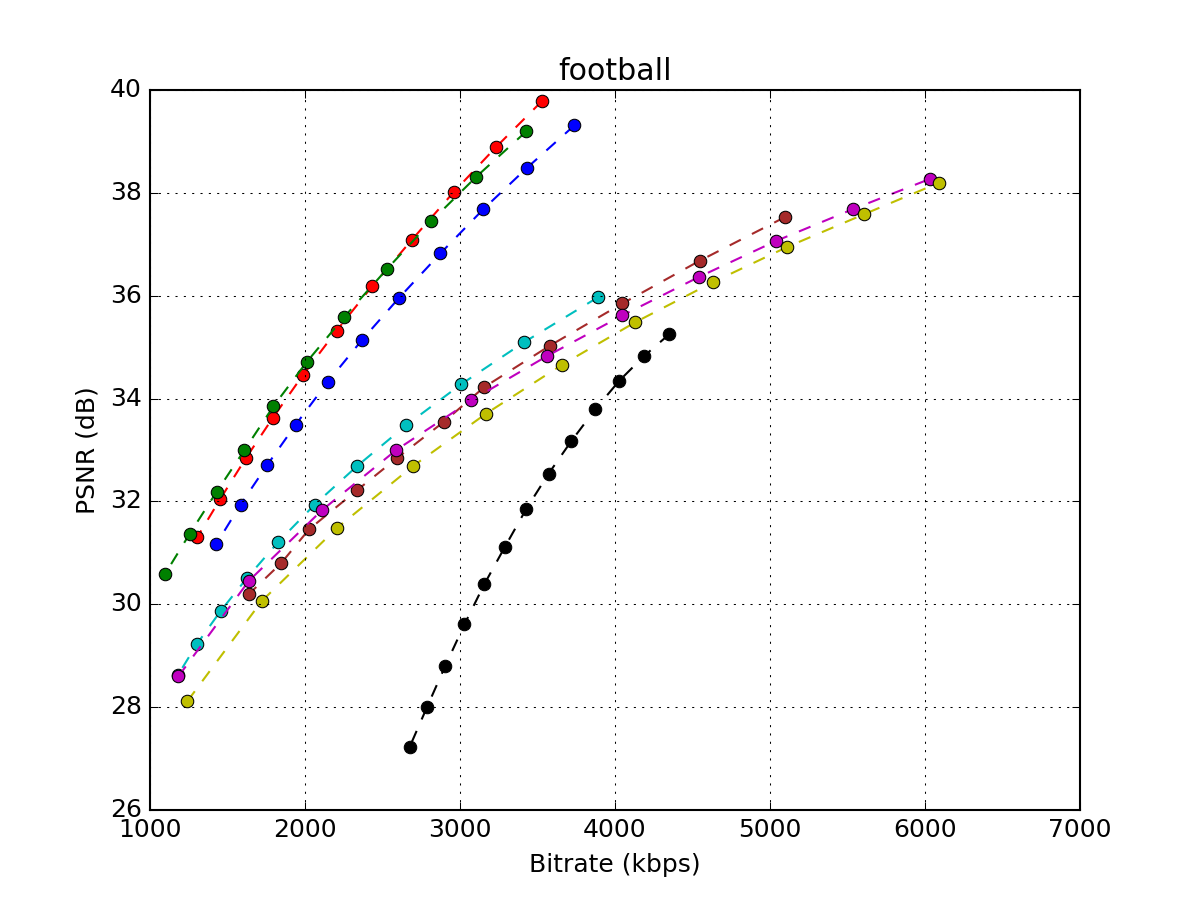}
}\quad
\subfloat     {
\includegraphics[width=.45\linewidth]{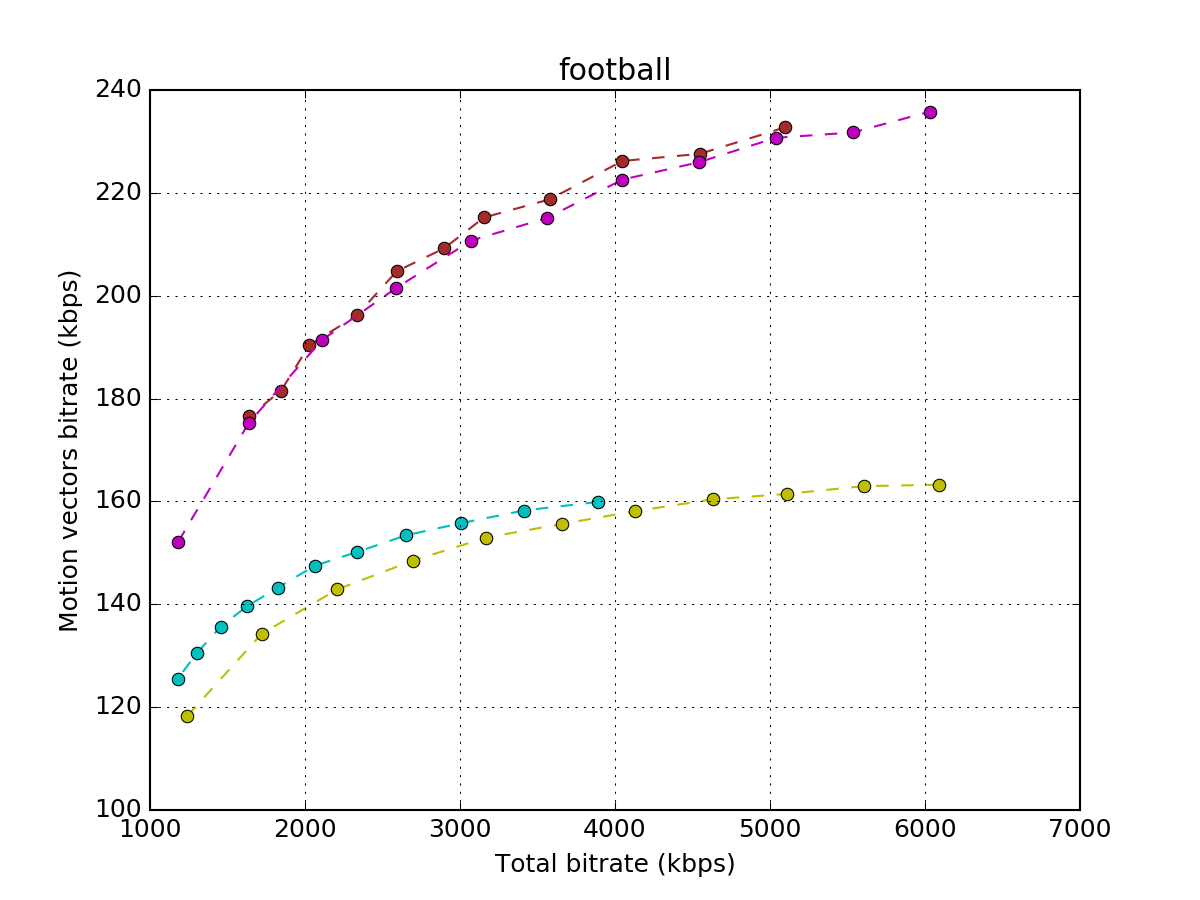}
}\\
\subfloat     {
\includegraphics[width=.45\linewidth]{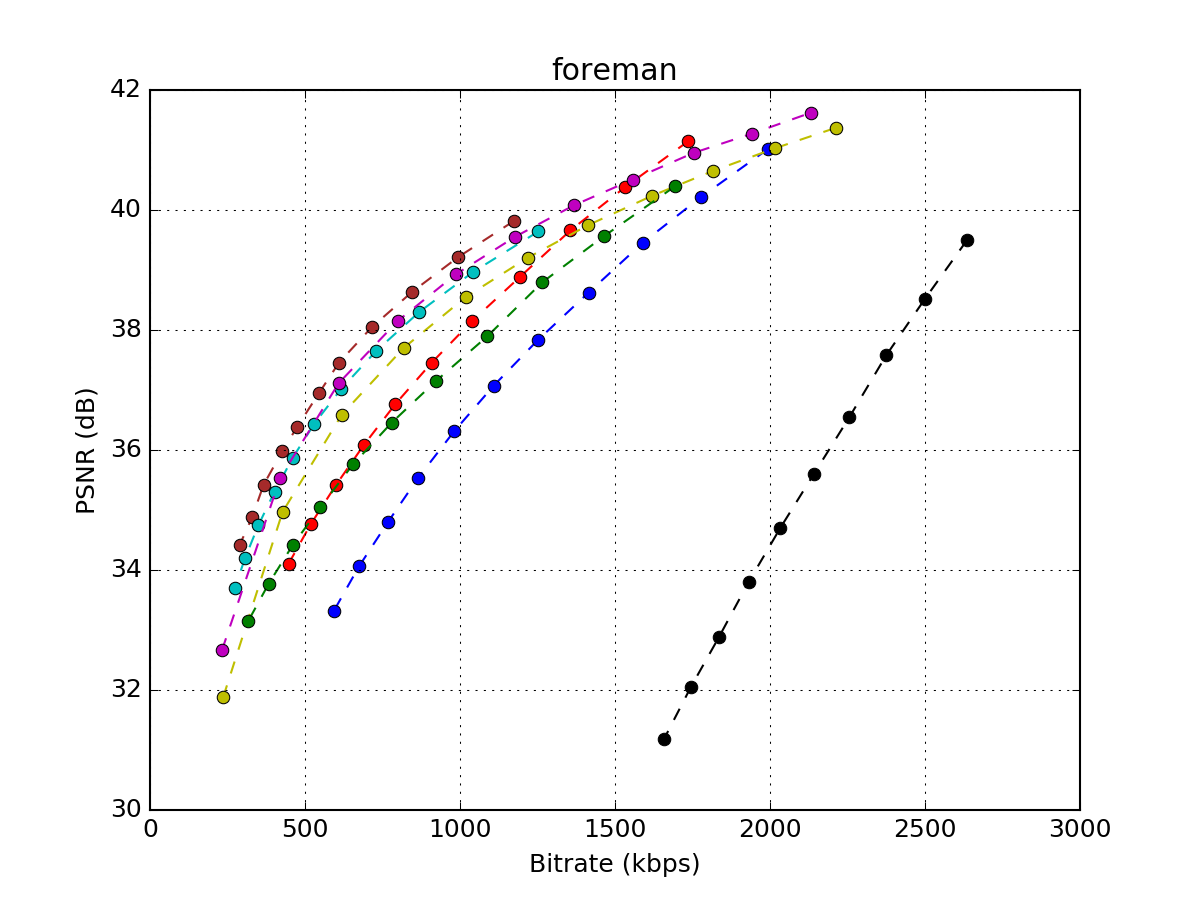}
}\quad
\subfloat     {
\includegraphics[width=.45\linewidth]{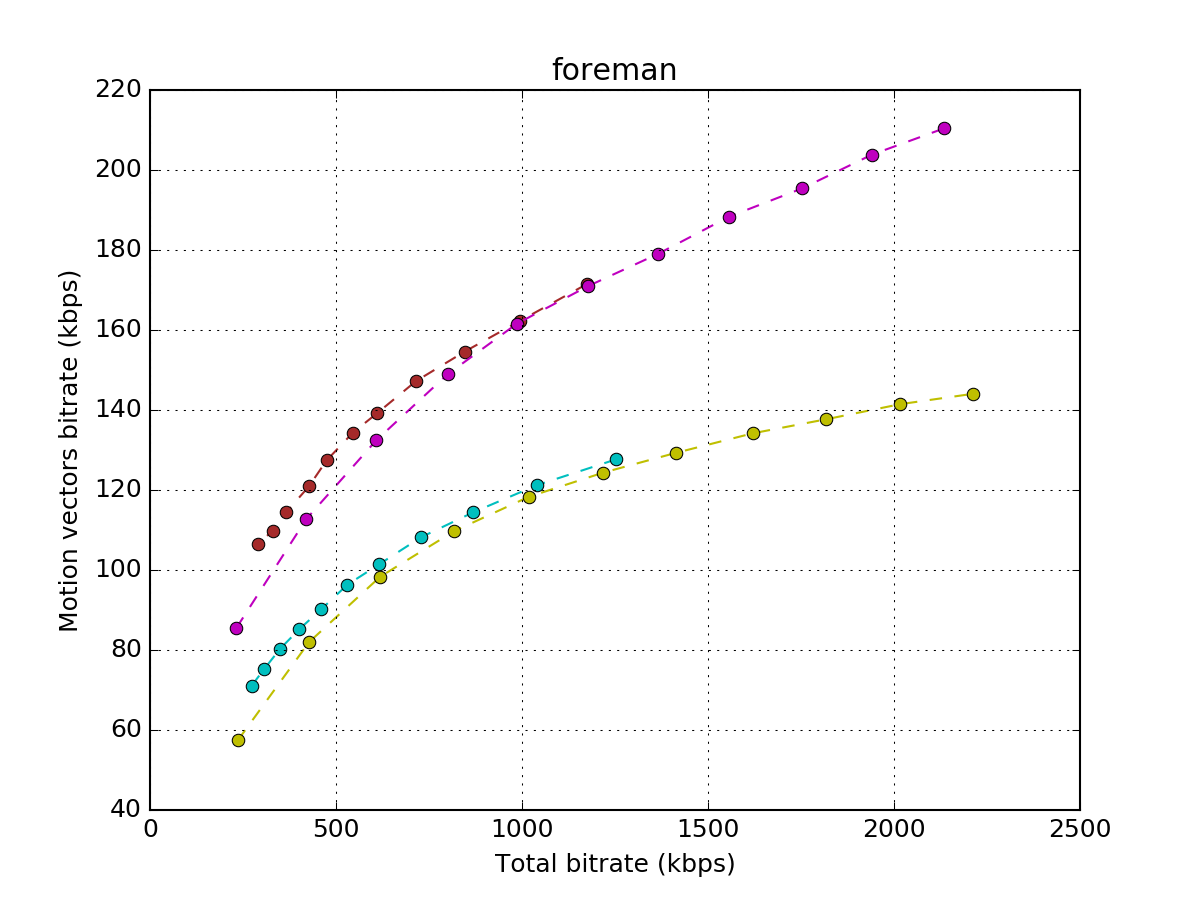}
}
\label{fig:compression1}
\end{center}
\end{figure}

\clearpage

\newcommand{\specialcell}[2][c]{%
  \begin{tabular}[#1]{@{}c@{}}#2\end{tabular}}

\begin{table}[h]
\caption{Bjontegaard delta PSNR metrics with respect to x264 SEQ-VBR (anchor). Positive values indicate higher performance compared to the anchor.}
\begin{tabular}{c|ccccccc}
              & \specialcell{LFP\\MSE} & \specialcell{LFP\\GAN} & \specialcell{HIE\\VBR} & \specialcell{HIE\\CBR} & \specialcell{SEQ\\CBR} & MC     & FD      \\ \hline
Coastguard    & 3.103   & 2.410   & 0.745   & 0.642   & -0.429  & 2.982  & -5.439  \\
Container     & 0.567   & -2.070  & 1.897   & 1.115   & -0.461  & -0.175 & -2.356  \\
Football      & 3.035   & 2.371   & -0.432  & -0.265  & -0.852  & 3.004  & -4.041  \\
Foreman       & -1.197  & -2.732  & 0.435   & 0.079   & -0.526  & -1.372 & -11.118 \\
Garden        & 3.739   & 3.150   & 1.168   & 0.711   & -0.996  & 3.021  & -9.173  \\
Hall monitor & -0.200  & -4.434  & 0.318   & -0.501  & -0.998  & -0.763 & -3.919  \\
Mobile        & 1.664   & 0.656   & 0.823   & 0.840   & -0.624  & -0.216 & -11.428 \\
Tennis        & -0.220  & -1.094  & 0.231   & -0.204  & -0.776  & 0.361  & -4.342 
\end{tabular}
\label{table:bjontegaard}
\end{table}


\chapter{Conclusion}

In this thesis we have firstly introduced a deep convolutional neural network for the task of video frame prediction. We have trained this model to minimize mean square loss and obtained accurate frame predictions.

We have also employed adversarial training and generated sharper and more realistic images. We have shown that, though GANs perform better qualitatively, the PSNR values of their outputs are lower. We have further demonstrated that adversarial training deteriorates compression performance, hence it is redundant for video compression.

Using our neural networks, we have built encoders and decoders for video compression. These encoders do not require the calculation or transmission of motion vectors, so the size of the transferred data is less compared to the standard codecs. Moreover, we have compared our method to x264 codec, a benchmark for video compression. We have outperformed the anchor method of sequential x264 with variable bitrate, in standard quantitative metrics, for more than half of the test videos. For half of the videos, we have outperformed the x264 codec with best possible configuration, namely hierarchical prediction with variable bitrate.

While the usage of deep neural networks is a relatively recent practice, it sets new standards in common engineering problems every day. There isn't a standard codec that employs a neural network yet, but it is very probable in the near future. To obtain better compression performance using neural networks, bidirectional prediction is a topic that should be investigated upon. Also the compression of the neural network itself to obtain a smaller network with shorter runtime can enable faster video compression. The creation of a standalone video compression library that uses neural networks can be the final step to create widespread usage in video compression with neural networks.

\bibliography{references}
\bibliographystyle{IEEEtran}

\end{document}